\documentclass[draftclsnofoot,peerreview,english]{IEEEtran}

\usepackage[T1]{fontenc}
\usepackage[latin9]{inputenc}
\usepackage{amsmath}
\usepackage{amssymb}
\usepackage{mathrsfs}
\usepackage{bbm}
\usepackage{mathtools}
\usepackage{empheq}
\usepackage{graphicx}
\usepackage{amsthm}	
\usepackage{hyperref}
\usepackage{cite}
\usepackage{accents}


\usepackage{babel}

\usepackage{graphicx}
\usepackage{wrapfig}
\usepackage{epsfig}
\usepackage{graphicx}
\usepackage{color}
\usepackage{mdframed}
\usepackage{mathrsfs} 
\usepackage{amsfonts}
\usepackage{latexsym}
\usepackage{amsmath}
\usepackage{amssymb}
\usepackage{color}
\usepackage{float} 
\usepackage{floatflt} 
\usepackage{caption}
\usepackage{subcaption}

\usepackage{pdfpages}

\newcommand{\ml}[1]{{\color{black} #1}}
\newcommand{\mikel}[1]{{\color{black} #1}}
\newcommand{\parham}[1]{{\color{black} #1}}

\newcommand{\suppress}[1]{}
\newcommand{\bx}{x^n}
\newcommand{\bxf}{x^{n_1}}
\newcommand{\bxs}{x^{n_2}}

\newcommand{\by}{y^n}
\newcommand{\byf}{y^{n_1}}
\newcommand{\bys}{y^{n_2}}

\newtheorem{theorem}{Theorem}[section]
\newtheorem{lemma}{Lemma}[section]
\newtheorem{cor}{Corollary}[section]

\newtheorem{claim}{Claim}[section]

\newtheorem{proposition}{Proposition}[section]

\def\cX{\mbox{$\cal{X}$}}
\def\cY{\mbox{$\cal{Y}$}}

\def\cN{\mbox{$\cal{N}$}}

\def\c{k}

\def\01{\{0,1\}}

\newcommand{\remove}[1]{}

\begin{document}

\title{Negligible Cooperation:
Contrasting the Maximal- and Average-Error Cases}

\author{Parham~Noorzad, Michael
Langberg, Michelle Effros%
\thanks{This material is based upon work supported by the 
National Science Foundation under Grant Numbers 1527524 and
1526771,
and has appeared in part in \cite{SingleBit, continuityConf}.}%
\thanks{P. Noorzad was with the California
Institute of Technology, Pasadena, CA 91125 USA.
He is now with Qualcomm Technologies, Inc., San Diego, CA 
92121 USA (email: parham@qti.qualcomm.com). }
\thanks{M. Langberg is with the State
University of New York at Buffalo, Buffalo, NY 14260 USA
(email: mikel@buffalo.edu).}
\thanks{M. Effros is with the California
Institute of Technology, Pasadena, CA 91125 USA
(email: effros@caltech.edu). }}
\maketitle

\begin{abstract}
In communication networks, cooperative strategies
are coding schemes where network nodes work together 
to improve network performance metrics such as the total rate delivered across the network.
This work studies \emph{encoder} cooperation
in the setting of a discrete multiple access channel (MAC) with
two encoders and a single decoder. A network node, here called
the cooperation facilitator (CF), that is connected to both 
encoders via rate-limited links, enables 
the cooperation strategy. Previous work by the authors 
presents two classes of MACs: (i) one class where the 
\emph{average-error} sum-capacity has an infinite derivative 
in the limit where CF output link capacities approach zero, and (ii)
a second class of MACs where
the \emph{maximal-error} sum-capacity is not continuous 
at the point where the output link capacities of the CF equal zero.
This work contrasts the power of the CF in the maximal- and average-error
cases, showing that a {\emph{constant number of bits} communicated} 
over the CF output link
can yield a positive gain in the maximal-error sum-capacity, while
a far greater number of bits, even numbers that grow sublinearly
in the blocklength, can never yield a non-negligible gain in the 
average-error sum-capacity.
\end{abstract}


\section{Introduction} \label{sec:intro}
Interference is an important limiting factor
in the capacities of many communication 
networks. One way to reduce interference is to enable
network nodes to work together to coordinate
their transmissions. Strategies that employ coordinated
transmissions are called cooperation strategies.

\begin{figure} 
	\begin{center}
		\includegraphics[scale=0.25]{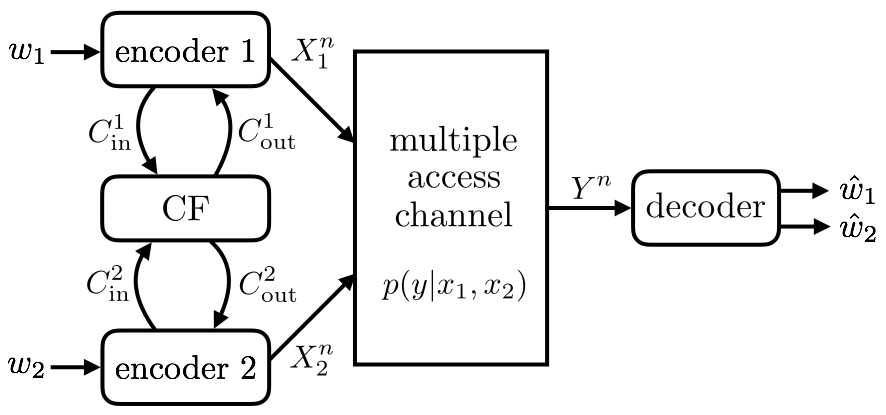}
		\caption{A network consisting of two encoders that 
		initially cooperate via a CF and then, based on the 
		information they receive, transmit their codewords
		over the MAC to the decoder.} \label{fig:model}
	\end{center}
\end{figure}

Perhaps the simplest cooperation strategy is 
``time-sharing'' (e.g., \cite[Theorem 15.3.2]{CoverThomas}), 
where nodes avoid interference by
taking turns transmitting. A popular alternative
model is the ``conferencing''
cooperation model \cite{WillemsMAC};
in conferencing, unlike in time-sharing, 
encoders share information about the messages they 
wish to transmit and use that shared information to 
coordinate their channel inputs. 
In this work, we employ a similar approach, but in
our cooperation model, encoders communicate indirectly.
Specifically, the encoders communicate through another
node, which we call the cooperation facilitator
(CF) \cite{reliability, kUserMAC}. Figure \ref{fig:model} 
depicts the CF model in the
two-user multiple access channel (MAC) scenario.

The CF enables cooperation between the encoders through
its rate-limited input and output links. Prior to choosing
a codeword to transmit over the channel, each encoder sends a
function of its message to the CF. The CF uses the information 
it receives from \emph{both} encoders to compute
a rate-limited function for each encoder. It then transmits the 
computed values over its output links. Finally, each encoder selects a
codeword using its message and the information it receives from
the CF.

To simplify our discussion in this section, suppose the CF input link capacities
both equal $C_\mathrm{in}$ and the CF output link capacities both
equal $C_\mathrm{out}$. If $C_\mathrm{in}\leq C_\mathrm{out}$, then
the optimal strategy for the CF is to simply
forward the information it receives from one encoder to the other.
Using the capacity region of the MAC with
conferencing encoders \cite{WillemsMAC}, it follows that the 
average-error sum-capacity gain of CF cooperation 
in this case is bounded from above by $2C_\mathrm{in}$ 
and does not depend on the precise value of $C_\mathrm{out}\geq C_\mathrm{in}$. 
If $C_\mathrm{in}>C_\mathrm{out}$, however, the situation 
is more complicated since the CF can no longer forward all of its
incoming information. While the $2C_\mathrm{in}$ upper bound is still
valid, the dependence of the sum-capacity gain on $C_\mathrm{out}$ is
less clear. If the CF simply forwards part of the information it 
receives, then again by \cite{WillemsMAC}, the average-error sum-capacity
gain is at most $2C_\mathrm{out}$. The $2C_\mathrm{out}$ bound has
an intuitive interpretation: it reflects the amount
of information the CF shares with the encoders, perhaps suggesting
that the benefit of sharing information with rate $2C_\mathrm{out}$ 
with the encoders is at most $2C_\mathrm{out}$. It turns out, though, 
that a much larger gain 
is possible through more sophisticated coding techniques. 
Specifically, in prior work \cite[Theorem 3]{kUserMAC}, we
show that for a class of MACs, for fixed $C_\mathrm{in}>0$,
the average-error sum-capacity has a derivative 
in $C_\mathrm{out}$ that is infinite at $C_\mathrm{out}=0$; that is, for small
$C_\mathrm{out}$, the gain resulting from cooperation exceeds any function
that goes through the origin and has bounded derivative. 

The large sum-capacity gain described above is not limited to 
the average-error scenario. In fact, in related work \cite[Proposition 5]{reliability}, 
we show that for any MAC for which, in the absence of cooperation, 
the average-error sum-capacity
is strictly greater than the maximal-error sum-capacity,
adding a CF and measuring  
the \emph{maximal-error} sum-capacity for fixed $C_\mathrm{in}>0$
gives a curve that is discontinuous at $C_\mathrm{out}=0$.
In this case, we say that ``negligible cooperation'' results in a 
non-negligible capacity benefit.

Given these earlier results, a number of important questions
remain open. For example, we wish to understand how many bits
from the CF are needed to achieve the discontinuities already 
shown to be possible in the maximal-error case. We also seek
to understand, in the average-error case, whether the 
sum-capacity gain can be discontinuous in $C_\mathrm{out}$.

For the first question, we note that while the demonstration
of discontinuity at $C_\mathrm{out}=0$ for the maximal-error
case proves that negligible cooperation can yield a non-negligible
benefit, it does not distinguish how many bits are required to
effect that change nor whether that number of bits must grow with
the blocklength $n$. We therefore begin by pushing that question
to its extreme: we seek to understand the minimal output rate from the
CF that can change network capacity. Our central result for the 
maximal-error case demonstrates that even a {constant number of bits} from the 
CF can yield a non-negligible impact on network capacity in
the maximal-error case.

For the second question, we seek to gain a similar understanding of 
how many bits from the CF are required to obtain a non-negligible 
change to network capacity in the average-error case. Since in this
case there are no prior results demonstrating the possibility of a
discontinuity in $C_\mathrm{out}$, we begin by \parham{investigating}
whether the sum-capacity in the average-error case can ever be 
discontinuous. Our central result for the average-error case is that
the average-error sum-capacity is continuous even at $C_\mathrm{out}=0$. 
(See Corollary \ref{cor:continuityAtZero}.) Our proof relies
on tools developed by Dueck \cite{DueckSC} to prove the strong 
converse for the MAC. Saeedi Bidokhti and Kramer \cite{SaeediKramer} and
Kosut and Kliewer\cite[Proposition 15]{KosutKliewer} also use Dueck's method
to address similar problems. Our application of Dueck's method first appears 
in \cite[Appendix C]{NoorzadPhD}.

In addition to the contributions above, our work also explicitly strengthens earlier results. Specifically,
we refer
the reader to Theorem \ref{thm:infSlope} in Section \ref{sec:knownApriori}
and Corollary \ref{cor:discontinuity} in Section \ref{sec:results}, which provide 
stronger versions of results derived in \cite{kUserMAC} and \cite{reliability},
respectively.

\section{Related work}
A continuity problem similar to the one considered here 
appears in studying rate-limited feedback over the MAC. In that setting, 
Sarwate and Gastpar \cite{SarwateGastpar} use the 
dependence-balance bounds of Hekstra and Willems
\cite{HekstraWillems} to show that as the feedback rate 
converges to zero, the average-error capacity region 
converges to the average-error capacity region of 
the same MAC in the absence of feedback. 

The problem we study here can also be formulated as an
``edge removal problem'' as introduced by Ho, Effros, and
Jalali \cite{HoEtAl, JalaliEtAl}. 
The edge removal problem seeks to quantify the capacity 
effect of removing a single edge from a network. 
While bounds on this capacity impact exist in a
number of limited scenarios (see, for example, \cite{HoEtAl} and \cite{JalaliEtAl}),
the problem remains open in the general case.
In the context of network coding, Langberg and Effros show that this problem is 
connected to a number of other open problems, including the difference between the 
$0$-error and $\epsilon$-error capacity regions \cite{LangbergEffros1}
and the difference between the lossless source coding regions for
independent and dependent sources \cite{LangbergEffros2}. 

In \cite{KosutKliewer}, Kosut and Kliewer present different variations
of the edge removal problem in a unified setting. In their terminology,
the present work investigates whether the network consisting of a MAC 
and a CF satisfies the ``weak edge removal property'' with respect 
to the average-error reliability criterion. A discussion in 
\cite[Chapter 1]{NoorzadPhD} summarizes the known results for 
each variation of the edge removal problem. 

The question of whether the capacity region of a network consisting
of noiseless links is continuous with respect to the link capacities
is studied by Gu, Effros, and Bakshi \cite{GuEffrosBakshi}
and Chan and Grant \cite{ChanGrant}. The present work differs from
\cite{GuEffrosBakshi, ChanGrant} in the network under consideration;
while our network does have noiseless links (the CF input and output 
links), it also contains a multiterminal component (the MAC) which 
may exhibit interference or noise; no such component appears in  
\cite{GuEffrosBakshi, ChanGrant}.

For the maximal-error case, our study focuses on the effect of 
a constant number of bits of communication in the memoryless setting.
For noisy networks {\em with memory}, it is not difficult 
to see that {even} one bit of communication may indeed affect the capacity region. 
For example, 
consider a binary symmetric channel 
whose error probability $\theta$ is chosen at random 
and then fixed for all time.  
If for $i\in\{1,2\}$, $\theta$ equals $\theta_i$
with positive probability $p_i$, and $0\leq\theta_1<\theta_2\leq1/2$,
then a single bit of feedback (not rate 1, but exactly one bit
no matter how large the blocklength) from the receiver to the transmitter 
 suffices to increase the capacity.
For memoryless channels, the question is far more 
subtle and is the subject of our study.

In the next section, we present the cooperation model we consider
in this work. 

\section{The Cooperation Facilitator Model} \label{sec:model}

In this work, we study cooperation between two encoders that 
communicate their messages to a decoder 
over a stationary, memoryless, and discrete MAC.
Such a MAC can be represented by the triple
\begin{equation*}
\big(\mathcal{X}_1\times\mathcal{X}_2,p(y|x_1,x_2),
\mathcal{Y}\big),
\end{equation*}
where $\mathcal{X}_1$, $\mathcal{X}_2$, and $\mathcal{Y}$ 
are finite sets and $p(y|x_1,x_2)$ is a 
conditional probability mass function. For any positive
integer $n\geq 2$, 
the $n$th extension of this MAC is given by
\begin{equation*}
  p(y^n|x_1^n,x_2^n)\coloneqq
  \prod_{t=1}^n p(y_t|x_{1t},x_{2t}).
\end{equation*}

For each positive integer $n$, called the
blocklength, and nonnegative real numbers $R_1$ and
$R_2$, called the rates, we next define a 
$(2^{nR_1},2^{nR_2},n)$-code
for communication over a MAC with a 
$(\mathbf{C}_\mathrm{in},\mathbf{C}_\mathrm{out})$-CF. Here 
$\mathbf{C}_\mathrm{in}=(C_\mathrm{in}^1,C_\mathrm{in}^2)$ and 
$\mathbf{C}_\mathrm{out}=(C_\mathrm{out}^1,C_\mathrm{out}^2)$
represent the capacities of the CF input and output links,
respectively. (See Figure \ref{fig:model}.)

\subsection{Positive Rate Cooperation}
For every $x\geq 1$, let $[x]$ 
denote the set $\{1,\dots,\lfloor x\rfloor\}$.
For $i\in\{1,2\}$, the transmission of encoder $i$ to 
the CF is represented by a mapping
\begin{equation*} 
  \varphi_i\colon [2^{nR_i}]
  \rightarrow [2^{nC_\mathrm{in}^i}].
\end{equation*}
The CF uses the information it receives from the 
encoders to compute a function
\begin{equation*} 
  \psi_i\colon [2^{nC_\mathrm{in}^1}]\times [2^{nC_\mathrm{in}^2}]
  \rightarrow [2^{nC_\mathrm{out}^i}]
\end{equation*}
for encoder $i$, where $i\in\{1,2\}$.
Encoder $i$ uses its message and what it
receives from the CF to select a codeword according to
\begin{equation*}
  f_i\colon [2^{nR_i}]\times [2^{nC_\mathrm{out}^i}]
  \rightarrow \mathcal{X}_i^n.
\end{equation*}
The decoder finds estimates of the transmitted messages
using the channel output. It is represented by a mapping
\begin{equation*}
  g\colon\mathcal{Y}^n\rightarrow
  [2^{nR_1}]\times [2^{nR_2}].
\end{equation*}
The collection of mappings 
\begin{equation*}
\big(\varphi_1,\varphi_2,\psi_1,\psi_2,f_1,f_2,g\big)
\end{equation*}
defines a $(2^{nR_1},2^{nR_2},n)$-code for the MAC
with a $(\mathbf{C}_\mathrm{in},\mathbf{C}_\mathrm{out})$-CF.\footnote{Technically,
the definition we present here is for a single round of cooperation.
As discussed in \cite{reliability}, it is possible to define cooperation 
via a CF over multiple rounds. However, this general scenario
does not alter our main proofs. This is due to the fact that in Lemma
\ref{lem:CsumAvgBounds}, the lower bound only needs one round of cooperation,
while the upper bound holds regardless of the number of rounds.}

\subsection{{Constant Size Cooperation}}
To address the setting of {a constant number of cooperation bits,} we modify the 
output link of the CF to have 
support {$[2^\c]$ for some fixed integer $\c$}; unlike the prior support $[2^{nC^i_\mathrm{out}}]$, the support of this link is independent of the blocklength $n$.
Then, for $i\in\{1,2\}$, the transmission of encoder $i$ to 
the CF is represented by a mapping
\begin{equation*} 
  \varphi_i\colon [2^{nR_i}]
  \rightarrow [2^{nC_\mathrm{in}^i}].
\end{equation*}
The CF uses the information it receives from the 
encoders to compute a function
\begin{equation*} 
  \psi_i\colon [2^{nC_\mathrm{in}^1}]\times [2^{nC_\mathrm{in}^2}]
  \rightarrow [2^\c]
\end{equation*}
for encoder $i$, where $i\in\{1,2\}$.
Encoder $i$, as before, uses its message and what it
receives from the CF to select a codeword according to
\begin{equation*}
  f_i\colon [2^{nR_i}]\times [2^\c]
  \rightarrow \mathcal{X}_i^n.
\end{equation*}
We now say that  
\begin{equation*}
\big(\varphi_1,\varphi_2,\psi_1,\psi_2,f_1,f_2,g\big)
\end{equation*}
defines a $(2^{nR_1},2^{nR_2},n)$-code for the MAC
with a $(\mathbf{C}_\mathrm{in},\frac{\c}{n})$-CF.

\subsection{Capacity Region}
For a fixed code, the probability of decoding
a particular transmitted message pair $(w_1,w_2)$ 
incorrectly equals
\begin{equation*}
  \lambda_n(w_1,w_2)\coloneqq
  \sum_{y^n\colon g(y^n)\neq (w_1,w_2)}
  p(y^n|f_1(w_1,z_1),f_2(w_2,z_2)),
\end{equation*}
where $z_1$ and $z_2$ are the CF outputs and 
are calculated, for $i\in\{1,2\}$, according to 
\begin{equation*}
  z_i = \psi_i\big(\varphi_1(w_1),\varphi_2(w_2)\big).
\end{equation*}
The \emph{average} probability of error is defined as 
\begin{equation*}
  P_{e,\mathrm{avg}}^{(n)}\coloneqq
  \frac{1}{2^{n(R_1+R_2)}}\sum_{w_1,w_2}
  \lambda_n(w_1,w_2),
\end{equation*}
and the \emph{maximal} probability of error is given by
\begin{equation*}
  P_{e,\mathrm{max}}^{(n)}\coloneqq
  \max_{w_1,w_2} \lambda_n(w_1,w_2).
\end{equation*}

A rate pair $(R_1,R_2)$ is achievable with respect to 
the average-error reliability criterion if there 
exists an infinite sequence of 
$(2^{nR_1},2^{nR_2},n)$-codes such that 
$P^{(n)}_{e,\mathrm{avg}}\rightarrow 0$ as 
$n\rightarrow\infty$. The average-error capacity region
of a MAC with a 
$(\mathbf{C}_\mathrm{in},\mathbf{C}_\mathrm{out})$-CF, denoted by 
$\mathscr{C}_\mathrm{avg}(\mathbf{C}_\mathrm{in},\mathbf{C}_\mathrm{out})$,
is the closure of the set of all rate pairs
that are achievable with respect to the average-error reliability
criterion. The average-error sum-capacity is defined as 
\begin{equation*}
  C_\mathrm{sum}(\mathbf{C}_\mathrm{in},\mathbf{C}_\mathrm{out})
  \coloneqq 
  \max_{(R_1,R_2)\in\mathscr{C}_\mathrm{avg}(\mathbf{C}_\mathrm{in},\mathbf{C}_\mathrm{out})} 
  (R_1+R_2).
\end{equation*}

By replacing $P^{(n)}_{e,\mathrm{avg}}$ with 
$P^{(n)}_{e,\mathrm{max}}$, we can similarly define
achievable rates with respect to the maximal-error reliability criterion, 
the maximal-error capacity region, and the maximal-error 
sum-capacity. For a MAC with a 
$(\mathbf{C}_\mathrm{in},\mathbf{C}_\mathrm{out})$-CF,
we denote the maximal-error capacity region 
and sum-capacity by 
$\mathscr{C}_\mathrm{max}(\mathbf{C}_\mathrm{in},\mathbf{C}_\mathrm{out})$
and
$C_\mathrm{sum,max}(\mathbf{C}_\mathrm{in},\mathbf{C}_\mathrm{out})$,
respectively.

\section{Prior Results on the
Sum-Capacity Gain of Cooperation} \label{sec:knownApriori}

We next review a number of results from 
\cite{reliability, kUserMAC} which describe the sum-capacity
gain of cooperation under the CF model. We begin with 
the average-error case.

Consider a discrete MAC 
$(\mathcal{X}_1\times\mathcal{X}_2,p(y|x_1,x_2),\mathcal{Y})$.
Let $p_\mathrm{ind}(x_1,x_2)=p_\mathrm{ind}(x_1)p_\mathrm{ind}(x_2)$ 
be a distribution that satisfies
\begin{equation} \label{eq:pind}
  I_\mathrm{ind}(X_1,X_2;Y)
  \coloneqq I(X_1,X_2;Y)\Big|_{p_\mathrm{ind}(x_1,x_2)}
  =\max_{p(x_1)p(x_2)}I(X_1,X_2;Y);
\end{equation}
subscript ``ind'' here denotes independence between the output
of encoders 1 and 2 in the absence of cooperation.
In addition, suppose that there exists a distribution 
$p_\mathrm{dep}(x_1,x_2)$ such that the support of $p_\mathrm{dep}(x_1,x_2)$
is contained in
the support of $p_\mathrm{ind}(x_1)p_\mathrm{ind}(x_2)$, 
\begin{equation*}
  I_\mathrm{dep}(X_1,X_2;Y)
  \coloneqq I(X_1,X_2;Y)\Big|_{p_\mathrm{dep}(x_1,x_2)},
\end{equation*}
and  
\begin{equation} \label{eq:pindApdep}
  I_\mathrm{dep}(X_1,X_2;Y)
  + D\big(p_\mathrm{dep}(y)\|p_\mathrm{ind}(y)\big)
  > I_\mathrm{ind}(X_1,X_2;Y);
\end{equation}
here $p_\mathrm{ind}(y)$ and $p_\mathrm{dep}(y)$
are the marginals on $\mathcal{Y}$ resulting from
$p_\mathrm{ind}(x_1,x_2)$ and $p_\mathrm{dep}(x_1,x_2)$,
respectively.
Let $\mathcal{C}^*$ denote the class of all discrete 
MACs for which input distributions $p_\mathrm{ind}$
and $p_\mathrm{dep}$, as described above, exist.  
Theorem \ref{thm:infSlope} below is a stronger version of 
\cite[Theorem 3]{kUserMAC} in the two-user case; the latter
result is stated as a corollary below.
A similar result holds for the Gaussian MAC 
\cite[Prop. 9]{kUserMAC}. 

\begin{theorem} \label{thm:infSlope}
Let 
$(\mathcal{X}_1\times\mathcal{X}_2,p(y|x_1,x_2),\mathcal{Y})$
be a MAC in $\mathcal{C}^*$, and suppose 
$(\mathbf{C}_\mathrm{in},\mathbf{v})
\in\mathbb{R}_{>0}^2\times \mathbb{R}_{>0}^2$.
Then there exists a constant $K>0$, which depends only on
the MAC and $(\mathbf{C}_\mathrm{in},\mathbf{v})$, such that when $h$
is sufficiently small,
\begin{equation} \label{eq:sqRootThm}
  C_\mathrm{sum}(\mathbf{C}_\mathrm{in},h\mathbf{v})
  -C_\mathrm{sum}(\mathbf{C}_\mathrm{in},\mathbf{0})
  \geq K\sqrt{h}+o(\sqrt{h}).
\end{equation}
\end{theorem}

The proof of Theorem \ref{thm:infSlope} appears in Subsection
\ref{subsec:infSlopeProof}.

In the above theorem, dividing both sides of (\ref{eq:sqRootThm})
by $h$ and letting $h\rightarrow 0^+$ results in the next corollary.\footnote{Note that
Corollary \ref{cor:infSlope} does not lead to any conclusions regarding continuity;
a function $f(x)$ with infinite derivative at $x=0$ can be continuous 
(e.g., $f(x)=\sqrt{x}$) or discontinuous 
(e.g., $f(x)=\lceil x\rceil)$.}

\begin{cor} \label{cor:infSlope}
For any MAC in $\mathcal{C}^*$ and any 
$(\mathbf{C}_\mathrm{in},\mathbf{v})
\in\mathbb{R}_{>0}^2\times \mathbb{R}_{>0}^2$,
\begin{equation*}
  \lim_{h\rightarrow 0^+}
  \frac{C_\mathrm{sum}(\mathbf{C}_\mathrm{in},h\mathbf{v})
  -C_\mathrm{sum}(\mathbf{C}_\mathrm{in},\mathbf{0})}{h}
  =\infty.
\end{equation*}
\end{cor}

We next describe the maximal-error sum-capacity gain.
While it is possible in the average-error scenario 
to achieve a sum-capacity that has an infinite slope, 
a stronger result is known in the maximal-error case. There exists a class
of MACs for which the maximal-error sum-capacity 
exhibits a discontinuity in the capacities of the
CF output links. This is stated formally in the next 
proposition, which is a special case of 
\cite[Proposition 5]{reliability}. The proposition relies on the
existence of a discrete MAC with average-error sum-capacity
larger than its maximal-error sum-capacity; that existence was 
first proven by Dueck \cite{Dueck}. We investigate further properties of
Dueck's MAC in \cite[Subsection VI-E]{reliability}.

\begin{proposition} \label{prop:discontinuity}
Consider a discrete MAC for which
\begin{equation} \label{eq:avgGeqMax}
  C_\mathrm{sum}(\mathbf{0},\mathbf{0})
  > C_\mathrm{sum,max}(\mathbf{0},\mathbf{0}).
\end{equation}
Fix $\mathbf{C}_\mathrm{in}\in\mathbb{R}^2_{>0}$.
Then 
$C_\mathrm{sum,max}(\mathbf{C}_\mathrm{in},
\mathbf{C}_\mathrm{out})$
is not continuous at 
$\mathbf{C}_\mathrm{out}=\mathbf{0}$. 
\end{proposition}

We next present the main results of this work.

\section{Our results: Continuity of Average- and Maximal-Error Sum-Capacities} 
\label{sec:results}

In the prior section, for a fixed $\mathbf{C}_\mathrm{in}$,
we discussed previous results regarding the value
of $C_\mathrm{sum}(\mathbf{C}_\mathrm{in},\mathbf{C}_\mathrm{out})$
and $C_\mathrm{sum,max}(\mathbf{C}_\mathrm{in},\mathbf{C}_\mathrm{out})$
as a function of $\mathbf{C}_\mathrm{out}$ at
$\mathbf{C}_\mathrm{out}=\mathbf{0}$. In this section, we do not
limit ourselves to the point $\mathbf{C}_\mathrm{out}=\mathbf{0}$;
rather, we study 
$C_\mathrm{sum}(\mathbf{C}_\mathrm{in},\mathbf{C}_\mathrm{out})$
over its entire domain. 
 
We begin by considering the case where the CF has full
access to the messages. Formally, for a given discrete MAC 
$(\mathcal{X}_1\times\mathcal{X}_2,p(y|x_1,x_2),\mathcal{Y})$,
let the components of 
$\mathbf{C}_\mathrm{in}^*=(C_\mathrm{in}^{*1},C_\mathrm{in}^{*2})$ 
be sufficiently large so that 
any CF with input link capacities $C_\mathrm{in}^{*1}$ and
$C_\mathrm{in}^{*2}$ has 
full knowledge of the encoders' messages. For example,
we can choose $\mathbf{C}_\mathrm{in}^*$ such that
\begin{equation*}
  \min\{C_\mathrm{in}^{*1},C_\mathrm{in}^{*2}\}
  >\max_{p(x_1,x_2)}I(X_1,X_2;Y).
\end{equation*}
Our first result addresses the continuity of 
$C_\mathrm{sum}(\mathbf{C}_\mathrm{in}^*,\mathbf{C}_\mathrm{out})$
as a function of $\mathbf{C}_\mathrm{out}$ over 
$\mathbb{R}^2_{\geq 0}$.

\begin{theorem} \label{thm:CsumAvgContinuity}
For any discrete MAC, the mapping
\begin{equation*}
  \mathbf{C}_\mathrm{out}
  \mapsto C_\mathrm{sum}(\mathbf{C}_\mathrm{in}^*,
  \mathbf{C}_\mathrm{out}),
\end{equation*}
defined on $\mathbb{R}^2_{\geq 0}$ is continuous.
\end{theorem}

While Theorem \ref{thm:CsumAvgContinuity} focuses on the scenario
where $\mathbf{C}_\mathrm{in}=\mathbf{C}_\mathrm{in}^*$,
the result is sufficiently strong to address the continuity
problem for a fixed, arbitrary $\mathbf{C}_\mathrm{in}$
at $\mathbf{C}_\mathrm{out}=\mathbf{0}$. 
To see this,
note that for all $\mathbf{C}_\mathrm{in}\in\mathbb{R}^2_{\geq 0}$, 
\begin{align} 
  C_\mathrm{sum}(\mathbf{C}_\mathrm{in},
  \mathbf{0}) &\leq
  C_\mathrm{sum}(\mathbf{C}_\mathrm{in},
  \mathbf{C}_\mathrm{out})\notag\\
  &\leq 
  C_\mathrm{sum}(\mathbf{C}_\mathrm{in}^*,
  \mathbf{C}_\mathrm{out}).\label{eq:corAtZeroProof}
\end{align}
Corollary \ref{cor:continuityAtZero}, below, now
follows from Theorem \ref{thm:CsumAvgContinuity}
by letting $\mathbf{C}_\mathrm{out}$ approach zero
in (\ref{eq:corAtZeroProof}) and noting that for all
$\mathbf{C}_\mathrm{in}\in\mathbb{R}^2_{\geq 0}$,
\begin{equation*}
  C_\mathrm{sum}(\mathbf{C}_\mathrm{in}^*,\mathbf{0})
  =C_\mathrm{sum}(\mathbf{C}_\mathrm{in},\mathbf{0})
  =C_\mathrm{sum}(\mathbf{0},\mathbf{0}).
\end{equation*}

\begin{cor} \label{cor:continuityAtZero}
For any discrete MAC and any fixed
$\mathbf{C}_\mathrm{in}\in\mathbb{R}^2_{\geq 0}$, 
the mapping
\begin{equation*} 
  \mathbf{C}_\mathrm{out}
  \mapsto C_\mathrm{sum}(\mathbf{C}_\mathrm{in},
  \mathbf{C}_\mathrm{out}),
\end{equation*}
is continuous at $\mathbf{C}_\mathrm{out}=\mathbf{0}$.
\end{cor}

Recall that Proposition \ref{prop:discontinuity} gives
a sufficient condition under which
$C_\mathrm{sum,max}(\mathbf{C}_\mathrm{in},\mathbf{C}_\mathrm{out})$
is not continuous at $\mathbf{C}_\mathrm{out}=\mathbf{0}$
for a fixed 
$\mathbf{C}_\mathrm{in}\in\mathbb{R}^2_{>0}$.
Corollary \ref{cor:continuityAtZero} implies that
the sufficient condition is also necessary. This is 
stated in the next corollary. We prove this corollary
in Subsection \ref{subsec:corDiscontinuityProof}.

\begin{cor} \label{cor:discontinuity}
Fix a discrete MAC and
$\mathbf{C}_\mathrm{in}\in\mathbb{R}^2_{>0}$.
Then 
$C_\mathrm{sum,max}(\mathbf{C}_\mathrm{in},
\mathbf{C}_\mathrm{out})$
is not continuous at 
$\mathbf{C}_\mathrm{out}=\mathbf{0}$ if and only if
\begin{equation}  \label{eq:avgBiggerThanMax}
  C_\mathrm{sum}(\mathbf{0},\mathbf{0})
  > C_\mathrm{sum,max}(\mathbf{0},\mathbf{0}).
\end{equation}
\end{cor}

We next describe the second main result of this paper.
Our first main result, Theorem \ref{thm:CsumAvgContinuity},
shows that $C_\mathrm{sum}(\mathbf{C}_\mathrm{in}^*,
\mathbf{C}_\mathrm{out})$ is continuous in 
$\mathbf{C}_\mathrm{out}$ over $\mathbb{R}^2_{\geq 0}$.
The next result shows that proving the continuity
of $C_\mathrm{sum}(\mathbf{C}_\mathrm{in},
\mathbf{C}_\mathrm{out})$ over 
$\mathbb{R}^2_{\geq 0}\times \mathbb{R}^2_{\geq 0}$
is equivalent to demonstrating its continuity on certain
axes. Specifically, it suffices to check the continuity
of $C_\mathrm{sum}$ when one of $C_\mathrm{out}^1$
and $C_\mathrm{out}^2$ approaches zero, while the
other arguments of $C_\mathrm{sum}$ are fixed
positive numbers. 

\begin{theorem} \label{thm:limitedCinContinuity}
For any discrete MAC, the mapping
\begin{equation*}
  (\mathbf{C}_\mathrm{in},\mathbf{C}_\mathrm{out})
  \mapsto C_\mathrm{sum}(\mathbf{C}_\mathrm{in},
  \mathbf{C}_\mathrm{out}),
\end{equation*}
defined on $\mathbb{R}^2_{\geq 0}\times\mathbb{R}^2_{\geq 0}$ is continuous
if and only if for all 
$(\mathbf{C}_\mathrm{in},\mathbf{C}_\mathrm{out})
\in\mathbb{R}^2_{>0}\times\mathbb{R}^2_{>0}$,  
we have
\begin{equation*}
  C_\mathrm{sum}\big(\mathbf{C}_\mathrm{in},
  (\tilde{C}_\mathrm{out}^1,C_\mathrm{out}^2)\big)
  \rightarrow 
  C_\mathrm{sum}\big(\mathbf{C}_\mathrm{in},
  (0,C_\mathrm{out}^2)\big)
\end{equation*}
as $\tilde{C}_\mathrm{out}^1\rightarrow 0^+$
and
\begin{equation*}
  C_\mathrm{sum}\big(\mathbf{C}_\mathrm{in},
  (C_\mathrm{out}^1,\tilde{C}_\mathrm{out}^2)\big)
  \rightarrow 
  C_\mathrm{sum}\big(\mathbf{C}_\mathrm{in},
  (C_\mathrm{out}^1,0)\big)
\end{equation*}
as $\tilde{C}_\mathrm{out}^2\rightarrow 0^+$.
\end{theorem}

We remark that using a time-sharing argument, 
it is possible to show that 
$C_\mathrm{sum}$ is concave on 
$\mathbb{R}^2_{\geq 0}\times\mathbb{R}^2_{\geq 0}$ and thus
continuous on its interior. Therefore, it suffices to study the 
continuity of $C_\mathrm{sum}$ on the boundary of 
$\mathbb{R}^2_{\geq 0}\times\mathbb{R}^2_{\geq 0}$. Note
that Theorem \ref{thm:limitedCinContinuity} leaves the continuity 
problem of the average-error sum-capacity open in one case. If
$C_\mathrm{in}^1$ and $C_\mathrm{in}^2$ are positive but not sufficiently large
and $C_\mathrm{out}^1>0$, then we have not established the continuity of the sum-capacity, solely
as function of $C_\mathrm{out}^2$, at $C_\mathrm{out}^2=0^+$. (Clearly, the symmetric 
case where $C_\mathrm{out}^2>0$ and $C_\mathrm{out}^1\rightarrow 0^+$ remains open as well.)
This last scenario remains a subject for future work.

Finally, we present the third main contribution of our work, 
the discontinuity of sum-capacity in the maximum-error setting 
when the outgoing edges of the CF can send only {a constant number of bits}.
\begin{theorem} \label{thm:1bit}
{For $\c=6$,} the discrete MAC presented in \cite{Dueck} satisfies 
\begin{equation} \label{eq:1bit}
  C_\mathrm{sum,max}(\mathbf{C}^*_\mathrm{in},
\c/n)
  > C_\mathrm{sum,max}(\mathbf{C}^*_\mathrm{in},\mathbf{0}).
\end{equation}
\end{theorem}

We prove our key results in the following sections.
In Sections~\ref{sec:fullKnowledgeCF},\ref{sec:arbitraryCF}, 
and \ref{sec:1bit}, we outline the proofs of Theorems \ref{thm:CsumAvgContinuity}, 
\ref{thm:limitedCinContinuity}, and \ref{thm:1bit}, respectively.
We provide detailed proofs of our claims in Section~\ref{sec:proofs}.

\section{Continuity of Sum-Capacity: The 
$\mathbf{C}_\mathrm{in}=\mathbf{C}_\mathrm{in}^*$
Case} \label{sec:fullKnowledgeCF}

We start our study of the continuity of $C_\mathrm{sum}(\mathbf{C}_\mathrm{in}^*,
\mathbf{C}_\mathrm{out})$ by
presenting lower and upper 
bounds in terms of an auxiliary function $\sigma(\delta)$ 
defined for $\delta\geq 0$
(Lemma \ref{lem:CsumAvgBounds}). 
This function is similar to a tool used by Dueck in \cite{DueckSC} but
differs with \cite{DueckSC} in its reliance on a
time-sharing random variable denoted by $U$.
The random variable $U$ plays two roles. First it ensures
that $\sigma$ is concave, which immediately proves the continuity of
$\sigma$ over $\mathbb{R}_{>0}$. Second, together with a 
lemma from \cite{DueckSC} (Lemma \ref{lem:Dueck} below), it helps us 
find a single-letter upper bound
for $\sigma$ (Corollary \ref{cor:Dueck}). We then use the 
single-letter upper bound to prove continuity at $\delta=0$. 

The following definitions are useful for the description of 
our lower and upper bounds for 
$C_\mathrm{sum}(\mathbf{C}_\mathrm{in}^*,
\mathbf{C}_\mathrm{out})$.
For every finite alphabet $\mathcal{U}$ and all $\delta\geq 0$, 
define the set of probability mass functions 
$\mathcal{P}^{(n)}_\mathcal{U}(\delta)$ on 
$\mathcal{U}\times\mathcal{X}_1^n\times\mathcal{X}_2^n$ as
\begin{equation*}
  \mathcal{P}^{(n)}_\mathcal{U}(\delta)
  :=\Big\{p(u,x_1^n,x_2^n)\Big |I(X_1^n;X_2^n|U)\leq n\delta\Big\}.
\end{equation*}
Intuitively, $\mathcal{P}^{(n)}_\mathcal{U}(\delta)$ captures a family
of ``mildly dependent'' input distributions for our MAC; this mild
dependence is parametrized by a bound $\delta$ on the per-symbol mutual information.
In the discussion that follows, we relate $\delta$ to the amount of information
that the CF shares with the encoders. For every positive integer $n$, let
$\sigma_n\colon\mathbb{R}_{\geq 0}\rightarrow\mathbb{R}_{\geq 0}$
denote the function\footnote{For $n=1$, this function also appears in 
the study of the MAC with negligible feedback \cite{SarwateGastpar}.} 
\begin{equation} \label{eq:fndelta}
  \sigma_n(\delta)\coloneqq\sup_{\mathcal{U}}\max_{p\in \mathcal{P}^{(n)}_\mathcal{U}(\delta)}
  \frac{1}{n}I(X_1^n,X_2^n;Y^n|U),
\end{equation}
where the supremum is over all finite sets $\mathcal{U}$. Thus 
$\sigma_n(\delta)$ captures something like the maximal sum-rate
achievable under the mild dependence described above. As we see in
Lemma \ref{lem:concavity}, conditioning on the random variable
$U$ in (\ref{eq:fndelta}) ensures that $\sigma_n$ is concave. 

For every $\delta\geq 0$, $(\sigma_n(\delta))_{n=1}^\infty$ satisfies 
a superadditivity property which appears in Lemma \ref{lem:superadditivity},
below. Intuitively, this property says that the sum-rate
of the best code of blocklength $m+n$ is bounded from below 
by the sum-rate of the concatenation of the best codes of
blocklengths $m$ and $n$. We prove this Lemma in Subsection 
\ref{subsec:superadditiveProof}.

\begin{lemma} \label{lem:superadditivity}
For all $m,n\geq 1$, all $\delta\geq 0$, and 
$\sigma_n(\delta)$ defined as in (\ref{eq:fndelta}), we have
\begin{equation*}
  (m+n)\sigma_{m+n}(\delta)\geq m\sigma_m(\delta)
  +n\sigma_n(\delta).
\end{equation*}
\end{lemma}

Given Lemma \ref{lem:superadditivity}, 
\cite[Appendix 4A, Lemma 2]{Gallager} now implies that
the sequence of mappings
$(\sigma_n)_{n=1}^\infty$ 
converges pointwise to some mapping
$\sigma\colon\mathbb{R}_{\geq 0}\rightarrow\mathbb{R}_{\geq 0}$,
and  
\begin{equation} \label{eq:FofDelta}
  \sigma(\delta) \coloneqq
  \lim_{n\rightarrow\infty}\sigma_n(\delta)
  =\sup_n \sigma_n(\delta).
\end{equation}

We next present our lower and upper bounds for 
$C_\mathrm{sum}(\mathbf{C}_\mathrm{in}^*,
\mathbf{C}_\mathrm{out})$ in terms of $\sigma$. The lower
bound follows directly from \cite[Corollary 8]{kUserMAC}.
We prove the upper bound in Subsection \ref{subsec:CsumAvgBounds}.

\begin{lemma} \label{lem:CsumAvgBounds}
For any discrete MAC and any $\mathbf{C}_\mathrm{out}\in\mathbb{R}^2_{\geq 0}$, 
we have
\begin{equation*}
  \sigma(C_\mathrm{out}^1+C_\mathrm{out}^2)
  -\min\{C_\mathrm{out}^1,C_\mathrm{out}^2\}
  \leq
  C_\mathrm{sum}(\mathbf{C}_\mathrm{in}^*,
  \mathbf{C}_\mathrm{out})
  \leq
  \sigma(C_\mathrm{out}^1+C_\mathrm{out}^2).
\end{equation*}
\end{lemma}
From the remark following Theorem \ref{thm:limitedCinContinuity},
we only need to prove that 
$C_\mathrm{sum}(\mathbf{C}_\mathrm{in}^*,\mathbf{C}_\mathrm{out})$
is continuous on the boundary of $\mathbb{R}^2_{\geq 0}$. 
On the boundary of $\mathbb{R}^2_{\geq 0}$, however, 
$\min\{C_\mathrm{out}^1,C_\mathrm{out}^2\}=0$. Thus it suffices
to show that $\sigma$ is continuous on $\mathbb{R}_{\geq 0}$,
which is stated in the next lemma. 

\begin{lemma} \label{lem:continuityOfF}
For any finite alphabet MAC, 
the function $\sigma$, defined by (\ref{eq:FofDelta}), 
is continuous on $\mathbb{R}_{\geq 0}$. 
\end{lemma}

To prove Lemma \ref{lem:continuityOfF}, we first consider 
the continuity of $\sigma$
on $\mathbb{R}_{>0}$ and then focus on the point $\delta=0$. 
Note that $\sigma$ is the pointwise limit
of the sequence of functions $(\sigma_n)_{n=1}^\infty$.
Lemma \ref{lem:concavity} uses a time-sharing argument as in 
\cite{CoverElGamalSalehi} to
show that each $\sigma_n$ is concave. (See Subsection 
\ref{subsec:concavity} for the proof.)
Therefore, $\sigma$ is concave as well, and
since $\mathbb{R}_{>0}$ is open, $\sigma$ is 
continuous on $\mathbb{R}_{>0}$.

\begin{lemma} \label{lem:concavity}
For all $n\geq 1$, $\sigma_n$ is concave 
on $\mathbb{R}_{\geq 0}$. 
\end{lemma}

To prove the continuity of $\sigma(\delta)$
at $\delta=0$, we find an upper bound
for $\sigma$ in terms of $\sigma_1$. For
some finite set $\mathcal{U}$ and $\delta>0$,
consider a distribution 
$p(u,x_1^n,x_2^n)\in\mathcal{P}^{(n)}_\mathcal{U}(\delta)$.
By the definition of $\mathcal{P}^{(n)}_\mathcal{U}(\delta)$,
\begin{equation} \label{eq:multiletterMIbound}
  I(X_1^n;X_2^n|U)\leq n\delta.
\end{equation}
Finding a bound for $\sigma$ in terms of $\sigma_1$ requires
a single-letter version of (\ref{eq:multiletterMIbound}). 
In \cite{DueckSC}, Dueck presents the necessary result. 
We present Dueck's result in the next lemma and provide
the proof in Subsection \ref{subsec:Dueck}.

\begin{lemma}[Dueck's Lemma \cite{DueckSC}] \label{lem:Dueck}
Fix positive reals $\epsilon$ and $\delta$, positive integer $n$, and finite
alphabet $\mathcal{U}$. If $p\in\mathcal{P}^{(n)}_\mathcal{U}(\delta)$, then
there exists a set $T\subseteq [n]$ satisfying 
$|T|\leq n\delta/\epsilon$
such that 
\begin{equation*}
  \forall\: t\notin T\colon
  I(X_{1t};X_{2t}|U,X_1^T,X_2^T)\leq\epsilon,
\end{equation*}
where for $i\in\{1,2\}$, $X_i^T\coloneqq (X_{it})_{t\in T}$.
\end{lemma}

Corollary \ref{cor:DueckWringing} uses Lemma \ref{lem:Dueck} to find an upper
bound for $\sigma$ in terms of $\sigma_1$. The proof of this
corollary, in Subsection \ref{subsec:CorDueck}, combines ideas from 
\cite{DueckSC} with results derived here. 

\begin{cor} \label{cor:DueckWringing}
For all $\epsilon,\delta >0$, we have
\begin{equation*}
  \sigma\big(\delta\big)\leq 
  \frac{\delta}{\epsilon}\log 
  |\mathcal{X}_1||\mathcal{X}_2|
  +\sigma_1(\epsilon).
\end{equation*}
\end{cor}

By Corollary \ref{cor:DueckWringing}, we have
\begin{equation*}
  \sigma(0)\leq \lim_{\delta\rightarrow 0^+}\sigma(\delta)
  \leq \sigma_1(\epsilon).
\end{equation*}
If we calculate the limit $\epsilon\rightarrow 0^+$, we get
\begin{equation*}
  \sigma(0)\leq \lim_{\delta\rightarrow 0^+}\sigma(\delta)
  \leq \lim_{\epsilon\rightarrow 0^+}\sigma_1(\epsilon).
\end{equation*}
Since $\sigma(0)=\sigma_1(0)$,\footnote{This follows from the
converse proof of the MAC capacity region in the 
absence of cooperation \cite[Theorem 15.3.1]{CoverThomas}.} it suffices to 
show that $\sigma_1(\delta)$ is continuous
at $\delta=0$. 
Recall that $\sigma_1$ is defined as
\begin{equation} \label{eq:sigmaOneDef}
  \sigma_1(\delta)\coloneqq
  \sup_\mathcal{U}
  \max_{p\in \mathcal{P}^{(1)}_\mathcal{U}(\delta)}
  I(X_1,X_2;Y|U).
\end{equation}
Since in (\ref{eq:sigmaOneDef}), the supremum is over \emph{all} finite sets 
$\mathcal{U}$, it is difficult to find an upper
bound for $\sigma_1(\delta)$ near $\delta=0$ directly. 
Instead we first show, in Subsection \ref{subsec:cardinalityProof}, 
that it is possible to assume that $\mathcal{U}$ 
has at most two elements. 
\begin{lemma}[Cardinality of $\mathcal{U}$] 
\label{lem:cardinality}
In the definition of $\sigma_1(\delta)$, 
it suffices to calculate the supremum over all sets $\mathcal{U}$
with $|\mathcal{U}|\leq 2$. 
\end{lemma}

In Subsection \ref{subsec:continuityFnProof},
we prove the continuity of $\sigma_1$ at $\delta=0$ 
from Lemma \ref{lem:cardinality} using standard tools,
such as Pinsker's inequality \cite[Lemma 17.3.3]{CoverThomas} and the 
$L_1$ lower bound of KL divergence \cite[Lemma 11.6.1]{CoverThomas}.
The continuity of $\sigma_1$ on $\mathbb{R}_{>0}$
follows from the concavity of $\sigma_1$ on $\mathbb{R}_{\geq 0}$.

\begin{lemma}[Continuity of $\sigma_1$] \label{lem:continuityFn}
The function $\sigma_1$ is continuous on $\mathbb{R}_{\geq 0}$.
\end{lemma}

\section{Continuity of Sum-Capacity:
Arbitrary $\mathbf{C}_\mathrm{in}$}
\label{sec:arbitraryCF}

In this section, we study the continuity of
$C_\mathrm{sum}(\mathbf{C}_\mathrm{in},\mathbf{C}_\mathrm{out})$
over $\mathbb{R}^2_{\geq 0}\times\mathbb{R}^2_{\geq 0}$
with the aim of proving Theorem
\ref{thm:limitedCinContinuity}. 

Fix 
$(\mathbf{C}_\mathrm{in},\mathbf{C}_\mathrm{out})$.
For arbitrary 
$(\mathbf{\tilde{C}}_\mathrm{in},\mathbf{\tilde{C}}_\mathrm{out})$,
the triangle inequality implies
\begin{align}
  \MoveEqLeft
  \big|
  C_\mathrm{sum}(\mathbf{\tilde{C}}_\mathrm{in},\mathbf{\tilde{C}}_\mathrm{out})
  -C_\mathrm{sum}(\mathbf{C}_\mathrm{in},\mathbf{C}_\mathrm{out})\big|\notag\\
  &\leq \big|
  C_\mathrm{sum}(\mathbf{\tilde{C}}_\mathrm{in},\mathbf{\tilde{C}}_\mathrm{out})
  -C_\mathrm{sum}(\mathbf{C}_\mathrm{in},\mathbf{\tilde{C}}_\mathrm{out})
  \big|+ \big|
  C_\mathrm{sum}(\mathbf{C}_\mathrm{in},\mathbf{\tilde{C}}_\mathrm{out})
  -C_\mathrm{sum}(\mathbf{C}_\mathrm{in},\mathbf{C}_\mathrm{out})\big|
  \label{eq:triangleIneqArbitCin}
\end{align}
We study this bound in the limit
$(\mathbf{\tilde{C}}_\mathrm{in},\mathbf{\tilde{C}}_\mathrm{out})
\rightarrow
(\mathbf{C}_\mathrm{in},\mathbf{C}_\mathrm{out})$. 
We begin by considering the first term in (\ref{eq:triangleIneqArbitCin}).

\begin{lemma}[Continuity of Sum-Capacity in $\mathbf{C}_\mathrm{in}$]
\label{lem:continuityCin}
There exists a function
\begin{equation*}
  \Delta\colon\mathbb{R}^2_{\geq 0}\times\mathbb{R}^2_{\geq 0}
  \rightarrow\mathbb{R}_{\geq 0}
\end{equation*}
that satisfies 
\begin{equation*}
  \lim_{\mathbf{\tilde{C}}_\mathrm{in}\rightarrow\mathbf{C}_\mathrm{in}}
  \Delta(\mathbf{C}_\mathrm{in},\mathbf{\tilde{C}}_\mathrm{in})=0,
\end{equation*}
and for any finite alphabet MAC and 
$(\mathbf{C}_\mathrm{in},\mathbf{\tilde{C}}_\mathrm{in},\mathbf{C}_\mathrm{out})
\in\mathbb{R}^2_{\geq 0}\times\mathbb{R}^2_{\geq 0}\times\mathbb{R}^2_{\geq 0}$,
we have
\begin{equation*}
  \big|C_\mathrm{sum}(\mathbf{C}_\mathrm{in},\mathbf{C}_\mathrm{out})
  -C_\mathrm{sum}(\mathbf{\tilde{C}}_\mathrm{in},\mathbf{C}_\mathrm{out})\big|
  \leq \Delta(\mathbf{C}_\mathrm{in},\mathbf{\tilde{C}}_\mathrm{in}).
\end{equation*}
\end{lemma}

We prove Lemma \ref{lem:continuityCin} in Subsection \ref{subsec:continuityCin}.

Applying Lemma \ref{lem:continuityCin} to (\ref{eq:triangleIneqArbitCin}), we get
\begin{equation*}
  \big|C_\mathrm{sum}(\mathbf{\tilde{C}}_\mathrm{in},\mathbf{\tilde{C}}_\mathrm{out})
  -C_\mathrm{sum}(\mathbf{C}_\mathrm{in},\mathbf{C}_\mathrm{out})\big|
  \leq \Delta(\mathbf{C}_\mathrm{in},\mathbf{\tilde{C}}_\mathrm{in})
  +\big|
  C_\mathrm{sum}(\mathbf{C}_\mathrm{in},\mathbf{\tilde{C}}_\mathrm{out})
  -C_\mathrm{sum}(\mathbf{C}_\mathrm{in},\mathbf{C}_\mathrm{out})\big|.
\end{equation*}
Thus to calculate the limit 
$(\mathbf{\tilde{C}}_\mathrm{in},\mathbf{\tilde{C}}_\mathrm{out})
\rightarrow (\mathbf{C}_\mathrm{in},\mathbf{C}_\mathrm{out})$, 
Lemma \ref{lem:continuityCout} investigates
\begin{equation*}
  \lim_{\mathbf{\tilde{C}}_\mathrm{out}
  \rightarrow \mathbf{C}_\mathrm{out}}
  \big|C_\mathrm{sum}(\mathbf{C}_\mathrm{in},\mathbf{\tilde{C}}_\mathrm{out})
  -C_\mathrm{sum}(\mathbf{C}_\mathrm{in},\mathbf{C}_\mathrm{out})\big|.
\end{equation*}
We prove this lemma in Subsection \ref{subsec:continuityCout}.

\begin{lemma}[Continuity of Sum-Capacity in $\mathbf{C}_\mathrm{out}$]
\label{lem:continuityCout}
For any finite alphabet MAC and 
$(\mathbf{C}_\mathrm{in},\mathbf{C}_\mathrm{out})
\in\mathbb{R}^2_{\geq 0}\times\mathbb{R}^2_{\geq 0}$,
proving that
\begin{equation*}
  \lim_{\mathbf{\tilde{C}}_\mathrm{out}\rightarrow\mathbf{C}_\mathrm{out}} 
  C_\mathrm{sum}(\mathbf{C}_\mathrm{in},\mathbf{\tilde{C}}_\mathrm{out})
  =C_\mathrm{sum}(\mathbf{C}_\mathrm{in},\mathbf{C}_\mathrm{out})
\end{equation*}
is equivalent to showing that 
for all  
$(\mathbf{C}_\mathrm{in},\mathbf{C}_\mathrm{out})
\in\mathbb{R}^2_{>0}\times\mathbb{R}^2_{>0}$, we have
\begin{align*}
  \lim_{\tilde{C}_\mathrm{out}^1\rightarrow 0^+}
  C_\mathrm{sum}\big(\mathbf{C}_\mathrm{in},
  (\tilde{C}_\mathrm{out}^1,C_\mathrm{out}^2)\big)
  &=C_\mathrm{sum}\big(\mathbf{C}_\mathrm{in},
  (0,C_\mathrm{out}^2)\big)\\
  \lim_{\tilde{C}_\mathrm{out}^2\rightarrow 0^+}
  C_\mathrm{sum}\big(\mathbf{C}_\mathrm{in},
  (C_\mathrm{out}^1,\tilde{C}_\mathrm{out}^2)\big)
  &=C_\mathrm{sum}\big(\mathbf{C}_\mathrm{in},
  (C_\mathrm{out}^1,0)\big).
\end{align*}
\end{lemma}

\section{Discontinuity of Sum-Capacity with a Constant Number of Cooperation Bits}
\label{sec:1bit}

In this section we prove Theorem~\ref{thm:1bit}.
We start by presenting Dueck's deterministic memoryless MAC from \cite{Dueck}.
Consider the MAC $\big(\mathcal{X}_1\times\mathcal{X}_2,p_\mathrm{Dueck}(y|x_1,x_2),\mathcal{Y}\big)$ 
where $\mathcal{X}_1 = \{a,b,A,B\}$, $\mathcal{X}_2 = \{0,1\}$, $\mathcal{Y}=\{a,b,c,A,B,C\}\times\{0,1\}$.
The probability transition matrix $p_\mathrm{Dueck}(y|x_1,x_2)=1$ where for
the deterministic mapping $W:\mathcal{X}_1\times\mathcal{X}_2\rightarrow\mathcal{Y}$, $y=W(x_1,x_2)$. 
The mapping $W$ is defined as
\[
W(x_1,x_2) \coloneqq \left\{
\begin{array}{ll}
(c,0) & \mbox{if $(x_1,x_2)\in\{(a,0),(b,0)\}$} \\
(C,1) & \mbox{if $(x_1,x_2)\in\{(A,1),(B,1)\}$} \\
(x_1,x_2) & \mbox{otherwise.}
\end{array}
\right.
\]
For positive integer $n$, we define the mapping 
$W^n:\mathcal{X}_1^n\times\mathcal{X}_2^n\rightarrow\mathcal{Y}^n$
as $(y_1^n,y_2^n)=W^n(x_1^n,x_2^n)$ if and only if for all $1\leq i\leq n$,
\begin{equation*}
  (y_{1i},y_{2i}) = W(x_{1i},x_{2i}).
\end{equation*}

Set $\mathbf{C}^*_\mathrm{in}=(\log{\mathcal{X}_1},\log{\mathcal{X}_2})=(2,1)$ 
to allow the CF to have access to both source messages $w_1$ and $w_2$.
We use the following theorem from \cite{Dueck}.
\begin{theorem}[Outer Bound on the Maximal-Error Sum-Capacity\cite{Dueck}]
\label{thm:Dueck}
For the MAC 
$\big(\mathcal{X}_1\times\mathcal{X}_2,p_\mathrm{Dueck}(y|x_1,x_2),\mathcal{Y}\big)$
defined above, we have
\begin{equation*}
C_\mathrm{sum,max}(\mathbf{0},\mathbf{0}) \leq 
\max_{0\leq p\leq 1/2} 
\Big[H(1/3)+2/3-p+H(p)\Big].
\end{equation*}
\end{theorem}

Optimizing over $p$, and noting that 
$C_\mathrm{sum,max}(\mathbf{0},\mathbf{0})=C_\mathrm{sum,max}(\mathbf{C}^*_\mathrm{in},\mathbf{0})$, 
Theorem~\ref{thm:Dueck} directly implies the following corollary.
\begin{cor}
\label{cor:Dueck}
For the MAC 
$\big(\mathcal{X}_1\times\mathcal{X}_2,p_\mathrm{Dueck}(y|x_1,x_2),\mathcal{Y}\big)$
and $p^*=1/3$, we have
\begin{equation*}
  C_\mathrm{sum,max}(\mathbf{C}^*_\mathrm{in},\mathbf{0}) \leq H(p^*)+H(1/3)+2/3-p^* \leq 2.1632.
\end{equation*}
\end{cor}

To conclude the proof of Theorem~\ref{thm:1bit}, we now show that for some integer $\c$, $C_\mathrm{sum,max}(\mathbf{C}^*_\mathrm{in},\c/n) > C_\mathrm{sum,max}(\mathbf{C}^*_\mathrm{in},\mathbf{0})$.
\mikel{Specifically, for any $\epsilon>0$, $\delta>0$, $\c= \lceil \log{2\lceil 3/\delta \rceil}\rceil+1$, and $n$ sufficiently large, we define a $(2^{nR_1},2^{nR_2},n)$-code for our MAC
with a $(\mathbf{C}^*_\mathrm{in},\c/n)$-CF with zero maximum error} in which $R_1=(1.5 - \delta)(1 -\epsilon)$ and $R_2=1-\epsilon$.
Corollary~\ref{cor:Dueck} then implies that for $\delta=1/4$, $\c=6$, and $\epsilon>0$ sufficiently small we have $C_\mathrm{sum,max}(\mathbf{C}^*_\mathrm{in},\c/n) > C_\mathrm{sum,max}(\mathbf{C}^*_\mathrm{in},\mathbf{0})$.
We specify our code by presenting the functions $\big(\varphi_1,\varphi_2,\psi_1,\psi_2,f_1,f_2,g\big)$.

The functions $\varphi_1$ and $\varphi_2$ are the identity functions.
Function $\psi_1$ is a constant function which always returns $0^\c$.
That is, for our construction, the CF only needs to \mikel{send the cooperation information to encoder 2.
Thus the function $f_1$ does not depend on $\psi_1$} and maps message $w_1$ to $x_1^n$ deterministically according to a codebook $\{x_1^n(w_1)\}_{w_1 \in [2^{nR_1}]}$ to be specified later.
Before we define the \mikel{codebook $\{x_1^n(w_1)\}_{w_1 \in [2^{nR_1}]}$, and functions} $\psi_2$, $f_2$, and $g$, we set some notation and definitions.

Recall that $\cY=\{a,b,c,A,B,C\} \times \{0,1\}$.
We therefore define $\cY=\cY_{1} \times \cY_{2}$, where $\cY_1= \{A,B,C,a,b,c\}$ and $\cY_2 = \{0,1\}$.
Accordingly, let $y^n_1=(y_{1;1},\dots,y_{1;n}) \in {\cY^n_{1}}$ and $\by_{2}=(y_{2;1},\dots,y_{2;n}) \in \cY^n_{2}$.
Given transmitted codewords $\bx_1$ and $\bx_2$, recall that $(\by_{1},\by_{2})=W^n(\bx_1,\bx_2)$, where $W^n$ is the blocklength-$n$ extension of $W$ defined above. 
We use notation $\by_{1}=W^n_1(\bx_1,\bx_2)$ and $\by_{2}=W^n_2(\bx_1,\bx_2)$ to describe $W^n$.
Formally, we use $W(x_1,x_2)=(W_1(x_1,x_2),W_2(x_1,x_2))$.
Note that $\by_{2}=W^n_2(\bx_1,\bx_2)=\bx_2$ for all $\bx_1 \in \cX_1^n$.

Our communication scheme is divided into two phases, the first of blocklegth $n_1=(1-\epsilon)n$, and the second of blocklegth $n_2=\epsilon n$, with $n=n_1+n_2$. Roughly speaking, after the first phase the decoder will be able to {\em list-decode} the message $(w_1,w_2)$, and after the second it will be able to determine the correct message from its list. We thus refine our notation, and define for $i=1,2$:
$\bxf_i=(x_{i;1},\dots,x_{i;n_1})$,
$\bxs_i=(x_{i;n_1+1},\dots,x_{i;n})$,
$\byf_i=(y_{i;1},\dots,y_{i;n_1})$, and
$\bys_i=(y_{i;n_1+1},\dots,y_{i;n})$.
Accordingly, we represent the two phases in the encoding functions $f_1$ and $f_2$ as $f_{1;1},f_{1;2}$ and $f_{2;1}, f_{2;2}$. Namely, for messages $w_1$ and $w_2$ we have $\bxf_1=f_{1;1}(w_1)$, $\bxs_1=f_{1;2}(w_1)$, $\bxf_2=f_{2;1}(w_2,\psi_2(w_1,w_2))$, and $\bxs_2=f_{2;2}(w_2,\psi_2(w_1,w_2))$.
We start by discussing the first phase of communication.

Consider a vector $\byf=(\byf_1,\byf_2)$. 
Let $\mathcal{E}(\byf_{1})=|\{i \mid \by_{1;i} \in \{c,C\}\}|$ be the 
number of symbols in $\byf_{1}$ that equal $c$ or $C$.
Note that both $c$ and $C$ are the result of an {\em erasure}, since the channel maps
both $a$ and $b$, or both $A$ and $B$, 
to the same output.\footnote{Since output $Y_1=c$ occurs only with output $Y_2=0$ and output $Y_1=C$ occurs only with output $Y_2=1$, we could, alternatively, denote both by the erasure symbol ``$E$'' without losing any information.}
Let $(W^{-1})^{n_1}(\byf_1,\byf_2)$ be the set of inputs $(\bxf_1,\bxf_2) \in \cX_1^{n_1} \times \cX_2^{n_1}$ for which $W^{n_1}(\bxf_1,\bxf_2)=(\byf_{1},\byf_2)$;
then from the definition of our channel it holds that 
$|(W^{-1})^{n_1}(\byf_{1},\byf_2)|=2^{\mathcal{E}(\byf_{1})}$ since each erasure symbol must have resulted from one of two possible input symbols.

We say that the pair $(\byf_{1},\byf_2)$ is {\em good} if $\mathcal{E}(\byf_{1}) \leq n_1/2$ or equivalently, $|(W^{-1})^{n_1}(\byf_{1},\byf_2)| \leq 2^{n_1/2}$; thus the first phase of a channel output is good if at most half of the input symbols are erased.
In our encoding scheme, we would like to guarantee that $(\byf_{1},\byf_2)$ is always good.
This is accomplished using the cooperation bits of $\psi_2$, or more specifically, using the first bit $\psi_{2;1}$ of $\psi_2$.
Note that since we are interested in designing a code with $R_2=1-\epsilon$, we may represent $w_2$ as a binary vector of length $n(1-\epsilon)$. With this representation in mind, we specify $\psi_{2;1}$. The remaining bits of $\psi_2$ will be defined later.
\[
\psi_{2;1}(w_1,w_2) = \left\{
\begin{array}{ll}
0 & \mbox{if $W^{n_1}(f_{1;1}(w_1),w_2)$ is good}, \\
1 & \mbox{otherwise.}
\end{array}
\right.
\]
Note that our MAC \ml{and $f_{1;1}$ are} deterministic, and thus the CF can compute $W^{n_1}(f_{1;1}(w_1),w_2)$.

We are now ready to define the first phase of encoding, namely $f_{1;1}$ and $f_{2;1}$.
We start with $f_{2;1}$ which will equal either $w_2$ or its element-wise complement $\bar{w}_2$.
\[
f_{2;1}(w_2,\psi_{2;1}(w_1,w_2)) = \left\{
\begin{array}{ll}
w_2 & \mbox{if $\psi_{2;1}(w_1,w_2)=0$}, \\
\bar{w}_2 & \mbox{otherwise.}
\end{array}
\right.
\]
Since $\psi_{2;1}(w_1,w_2)$ is 0 when $W^{n_1}(f_{1;1}(w_1),w_2)$ is good and 1 otherwise, $f_{2:1}$ flips the message $w_2$ if and only if  the unflipped channel input yields a bad channel output.
The following claim now guarantees that $\byf=(\byf_1,\byf_2)$ is good.

\begin{claim}
\label{claim:good}
For any $w_1$ and $w_2$, \ml{and any deterministic mapping $f_{1;1}$,} if $W^{n_1}(f_{1;1}(w_1),w_2)$ is bad then $W^{n_1}(f_{1:1}(w_1),\bar{w}_2)$ is good. 
\end{claim}

\noindent
{\em Proof:} see Section~\ref{sec:claim:good}.

We now define \ml{$f_{1;1}$, i.e.,} the codebook $\{x_1^{n_1}(w_1)\}_{w_1 \in [2^{nR_1}]}$.
Consider first choosing the codebook uniformly at random from all subsets of size $2^{nR_1}$ of \mikel{$\cX_1^{n_1}$.}
Namely, let the function $f_{1;1}$ be distributed uniformly over all \ml{injective} functions $[2^{nR_1}] \rightarrow \cX_1^{n_1}$.
We show that with high probability over $f_{1:1}$, for any good received word $(\byf_{1},\byf_{2})$, there are at most $2\lceil 3/\delta \rceil$ codeword pairs $(\bxf_1,\bxf_2)$  that satisfy $W^{n_1}(\bxf_1,\bxf_2)=(\byf_{1},\byf_{2})$, where $\delta$ is defined in our choice of $R_1$ and is independent of the blocklength $n_1$.

\begin{claim}
\label{claim:good2}
For any sufficiently large $n_1$, with probability at least $1-2^{-0.4n_1}$ over $f_{1;1}$, for any good pair $(\byf_{1},\byf_2)$, there are at most $2\lceil 3/\delta \rceil$ message pairs $(w_1,w_2)$ such that $W^{n_1}(f_{1;1}(w_1),f_{2;1}(w_2,\psi_{2;1}(w_1,w_2)))=(\byf_{1},\byf_2)$. 
\end{claim}

\noindent
{\em Proof:} see Section~\ref{sec:claim:good2}.

\ml{Consider any function $f_{1;1}$  that satisfies the conditions of Claim~\ref{claim:good2}. 
Function $f_{1;1}$ defines the codebook  $\{x_1^{n_1}(w_1)\}_{w_1 \in [2^{nR_1}]}$ by setting $\bxf_1(w_1)=f_{1;1}(w_1)$.}
By Claim~\ref{claim:good} and our definition of $\psi_{2;1}(w_1,w_2)$, for any message pair $(w_1,w_2)$,
$W^{n_1}(f_{1;1}(w_1),f_{2;1}(w_2,\psi_{2;1}(w_1,w_2)))=(\byf_1,\byf_2)$ is always good.
Consider a preliminary decoding function $g_\mathrm{list}$ that, given $\byf=(\byf_1,\byf_2)$, returns all possible $(w_1,w_2)$ such that $W^{n_1}(f_{1;1}(w_1),f_{2;1}(w_2,\psi_{2;1}(w_1,w_2)))=(\byf_1,\byf_2)$.
By Claim~\ref{claim:good2}, the decoder $g_\mathrm{list}$ is a {\em list-decoder} with list size $2\lceil 3/\delta \rceil$.
Here, a list-decoder is a decoding function that returns a list of potential messages (of a limited size) which, under our deterministic channel model and code design, is guaranteed to  include the original source messages. 
%

We thus conclude that for $R_1=(1.5 - \delta)(1-\epsilon)$ and $R_2=1-\epsilon$, after the first phase of communication, the decoder can recover a list $L$ of size $\ell=2\lceil 3/\delta \rceil$ of potential message pairs that includes the original message pair $(w_1,w_2)$ of the encoder. Moreover, we note, due to the deterministic nature of our channel, that the CF can calculate the list $L$ and the location (in lexicographic order) of the original message $(w_1,w_2)$ in the list. 
As the list size is $\ell$, $\lceil\log{\ell}\rceil=\lceil\log{2\lceil 3/\delta \rceil}\rceil$ bits suffice to specify the location of the original message $(w_1,w_2)$ in the list.
We therefore define the remaining bits of $\psi_{2}(w_1,w_2)$, denoted by $\psi_{2;2}(w_1,w_2)$, to be the binary representation of the location of the original message $(w_1,w_2)$ in the list $L$ decoded by the decoder. We now show how encoders 1 and 2 can send $\psi_{2;2}(w_1,w_2)$ to the decoder in the second phase of our communication scheme, allowing it to recover the original message $(w_1,w_2)$ from the list $L$ obtained after the first phase.

To send $\psi_{2;2}(w_1,w_2)$ to the decoder during the second phase of communication (of blocklegth $n_2=\epsilon n$) we use the fact that the rate pair $(0,1)$ is in the zero error capacity region of our MAC.
This follows by noticing that $W_2(x_1,x_2)=x_2$.
Thus, we set $f_{1:2}(w_1)$ to be constant and $f_{2:2}(w_2,\psi_2(w_1,w_2))$ to equal $\psi_{2;2}(w_1,w_2)$ padded on the left by zeros (here, we use the fact that for sufficiently large $n$ it holds that $n_2=\epsilon n \geq \lceil\log{2\lceil 3/\delta \rceil}\rceil$). The decoder can now recover $\psi_{2;2}(w_1,w_2)$ and in turn the original message pair $(w_1,w_2)$ from its list. This concludes the proof of Theorem~\ref{thm:1bit}.

\section{Proofs}  \label{sec:proofs}

In this section, we begin with 
the proof of Corollary \ref{cor:discontinuity}.
We then provide detailed proofs of the 
lemmas appearing in Sections \ref{sec:fullKnowledgeCF},
\ref{sec:arbitraryCF}, and \ref{sec:1bit}.

\subsection{Proof of Theorem \ref{thm:infSlope}
\label{subsec:infSlopeProof}}

We seek to prove that for any MAC where cooperation increases sum-capacity
(formally specified by the definition of $\mathcal{C}^*$ in Section \ref{sec:knownApriori}), 
the benefit of cooperation grows at least as fast as the square root of $C_\mathrm{out}$.
 
Since $\mathbf{C}_\mathrm{in}\in\mathbb{R}^2_{>0}$, there 
exists $\lambda=\lambda(\mathbf{C}_\mathrm{in})\in (0,1)$ such that for $i\in\{1,2\}$, 
\begin{equation*}
  C_\mathrm{in}^i > \lambda C_\mathrm{in}^{*i}.
\end{equation*}
Using a time-sharing argument, it follows that 
\begin{equation} \label{eq:timeSharingLowerBound}
  C_\mathrm{sum}(\mathbf{C}_\mathrm{in},h\mathbf{v})
  \geq C_\mathrm{sum}(\lambda\mathbf{C}_\mathrm{in}^*,\lambda h\mathbf{v})
  \geq \lambda C_\mathrm{sum}(\mathbf{C}_\mathrm{in}^*,h\mathbf{v})
  +(1-\lambda) C_\mathrm{sum}(\mathbf{0},\mathbf{0}).
\end{equation}
Applying in (\ref{eq:timeSharingLowerBound}) the fact that
\begin{equation*}
  C_\mathrm{sum}(\mathbf{C}_\mathrm{in}^*,\mathbf{0})
  = C_\mathrm{sum}(\mathbf{C}_\mathrm{in},\mathbf{0})
  = C_\mathrm{sum}(\mathbf{0},\mathbf{0})
\end{equation*}
results in
\begin{equation} \label{eq:CinCinStarBound}
  C_\mathrm{sum}(\mathbf{C}_\mathrm{in},h\mathbf{v})
  - C_\mathrm{sum}(\mathbf{C}_\mathrm{in},\mathbf{0})
  \geq \lambda\big(C_\mathrm{sum}(\mathbf{C}_\mathrm{in}^*,h\mathbf{v})
  -C_\mathrm{sum}(\mathbf{C}_\mathrm{in}^*,\mathbf{0})\big).
\end{equation}
On the other hand, setting $\mathbf{C}_\mathrm{out}=h\mathbf{v}=h(v_1,v_2)$
in the lower bound of Lemma \ref{lem:CsumAvgBounds} gives
\begin{align}
  C_\mathrm{sum}(\mathbf{C}_\mathrm{in}^*,h\mathbf{v})
  &\geq \sigma\big(h(v_1+v_2)\big)-\min\{hv_1,hv_2\}\notag\\
  &\geq \sigma_1\big(h(v_1+v_2)\big)-\min\{hv_1,hv_2\},
  \label{eq:sigmaInTermsOfsigma1}
\end{align}
where (\ref{eq:sigmaInTermsOfsigma1}) follows by (\ref{eq:FofDelta}).
Note that 
\begin{equation*}
  \sigma_1(0) = C_\mathrm{sum}(\mathbf{0},\mathbf{0})
  = C_\mathrm{sum}(\mathbf{C}_\mathrm{in}^*,\mathbf{0}).
\end{equation*}
Therefore, by combining (\ref{eq:CinCinStarBound}) 
and (\ref{eq:sigmaInTermsOfsigma1}), we get
\begin{equation} \label{eq:CsumInTermsOfSigma1}
  C_\mathrm{sum}(\mathbf{C}_\mathrm{in},h\mathbf{v})
  - C_\mathrm{sum}(\mathbf{C}_\mathrm{in},\mathbf{0})
  \geq \lambda\big(\sigma_1(hv_1+hv_2)-\sigma_1(0)\big)
  -\lambda h\min\{v_1,v_2\}.
\end{equation}
Given (\ref{eq:CsumInTermsOfSigma1}), the following lemma
completes the proof.

\begin{lemma} \label{lem:sigmaOneLowerBound}
For any discrete memoryless MAC in $\mathcal{C}^*$, there exists a constant 
$K>0$, such that
\begin{equation*}
  \sigma_1(\delta)\geq \sigma_1(0)+K\sqrt{\delta}+o(\sqrt{\delta}). 
\end{equation*}
\end{lemma}

\textbf{Remark.} The key result in proving Lemma \ref{lem:sigmaOneLowerBound}
is the asymptotic equivalence of the Kullback-Liebler (KL) divergence
and the chi-squared divergence \cite[Theorem 4.1, p. 448]{CsiszarShields}.
Precisely, let $p(x)$ and $q(x)$ be distributions on a finite alphabet
$\mathcal{X}$. Then the KL and the chi-squared divergences between $p$
and $q$ are given by
\begin{align*}
  D\big(p(x)\|q(x)\big)
  &\coloneqq \sum_{x\in\mathcal{X}} p(x)\log\frac{p(x)}{q(x)}\\
  \chi^2\big(p(x),q(x)\big)
  &\coloneqq \sum_{x\in\mathcal{X}}\frac{\big(p(x)-q(x)\big)^2}{q(x)},
\end{align*}
respectively. Suppose $q$ has full support; that is, for all
$x\in\mathcal{X}$, $q(x)>0$. Then if $p$ tends to $q$ pointwise
on $\mathcal{X}$, then
\begin{equation*}
  \frac{D\big(p(x)\|q(x)\big)}{\chi^2\big(p(x),q(x)\big)}
  \rightarrow\frac{1}{2\ln 2}.
\end{equation*}
We remark that in a different setting,
the authors of \cite{WangEtAl} also use this property of 
the KL divergence to obtain a ``square-root law'' similar to 
Lemma \ref{lem:sigmaOneLowerBound} above.

\textit{Proof of Lemma \ref{lem:sigmaOneLowerBound}:}
Since our MAC is in $\mathcal{C^*}$, 
by definition, there
exists a distribution $p_0(x_1)p_0(x_2)$ with support 
$\mathcal{S}_0\subseteq\mathcal{X}_1\times\mathcal{X}_2$ that satisfies
\begin{equation*}
  I_0(X_1,X_2;Y) = \max_{p(x_1)p(x_2)}I(X_1,X_2;Y)
\end{equation*}
and a distribution $p_1(x_1,x_2)$ with support $\mathcal{S}_1\subseteq\mathcal{S}_0$
that satisfies 
\begin{equation*}
  I_1(X_1,X_2;Y) + D\big(p_1(y)\|p_0(y)\big) > I_0(X_1,X_2;Y).
\end{equation*}
For each $\lambda\in (0,1)$, define the distribution $p_\lambda(x_1,x_2)$ as
\begin{equation*}
  p_\lambda(x_1,x_2)\coloneqq
  (1-\lambda)p_0(x_1)p_0(x_2)+\lambda p_1(x_1,x_2).
\end{equation*}
We have 
\begin{align*}
  I_\lambda(X_1;X_2)
  &= \sum_{x_1,x_2}p_\lambda(x_1,x_2)
  \log\frac{p_\lambda(x_1,x_2)}{p_\lambda(x_1)p_\lambda(x_2)}\\
  &= \sum_{x_1,x_2}p_\lambda(x_1,x_2)
  \log\frac{p_\lambda(x_1,x_2)}{p_0(x_1)p_0(x_2)}
  + \sum_{x_1,x_2}p_\lambda(x_1,x_2)
  \log\frac{p_0(x_1)p_0(x_2)}{p_\lambda(x_1)p_\lambda(x_2)}\\
  &= \sum_{x_1,x_2}p_\lambda(x_1,x_2)
  \log\frac{p_\lambda(x_1,x_2)}{p_0(x_1)p_0(x_2)}
  + \sum_{x_1}p_\lambda(x_1)
  \log\frac{p_0(x_1)}{p_\lambda(x_1)}
  +\sum_{x_2}p_\lambda(x_2)
  \log\frac{p_0(x_2)}{p_\lambda(x_2)}\\
  &= D\big(p_\lambda(x_1,x_2)\|p_0(x_1)p_0(x_2)\big)
  -D\big(p_\lambda(x_1)\|p_0(x_1)\big)
  -D\big(p_\lambda(x_2)\|p_0(x_2)\big)
\end{align*}
Note $\mathcal{S}_\lambda$, the support of $p_\lambda(x_1,x_2)$, satisfies
\begin{equation} \label{eq:supportCondition}
  \mathcal{S}_\lambda\subseteq
  \mathcal{S}_0\cup\mathcal{S}_1
  \subseteq \mathcal{S}_0.
\end{equation}
Therefore, by \cite[Theorem 4.1, p. 448]{CsiszarShields} (described 
in the remark above), we have
\begin{equation} \label{eq:miK1}
  \lim_{\lambda\rightarrow 0^+}
  \frac{1}{\lambda^2}
  I_\lambda(X_1;X_2)
  = K_1,
\end{equation}
where 
\begin{equation} \label{eq:defK1}
  K_1\coloneqq \frac{1}{2\ln 2}
  \Big(\chi^2\big(p_1(x_1,x_2),p_0(x_1)p_0(x_2)\big)
  -\chi^2\big(p_1(x_1),p_0(x_1)\big)
  -\chi^2\big(p_1(x_2),p_0(x_2)\big)\Big).
\end{equation}
Since $\mathcal{S}_1\subseteq\mathcal{S}_0$, all chi-squared distances
in (\ref{eq:defK1}) are well-defined. Also, since mutual information 
is always nonnegative, by (\ref{eq:miK1}) we have $K_1\geq 0$. Fix
$\epsilon>0$, and define the mapping $\delta\colon [0,1]\rightarrow\mathbb{R}$
as
\begin{equation*}
  \delta(\lambda)\coloneqq 
  I_\lambda(X_1;X_2)+\epsilon\lambda^2.
\end{equation*}
Note that by (\ref{eq:miK1}), 
\begin{equation*}
  \lim_{\delta\rightarrow 0^+}
  \frac{\delta(\lambda)}{\lambda^2}
  = K_1+\epsilon>0.
\end{equation*}
Furthermore, since $\delta$ is continuously differentiable in $\lambda$, by the inverse
function theorem, there exists a function $\lambda^*\colon [0,\delta_0)\rightarrow [0,1]$
for some $\delta_0>0$ that satisfies
\begin{equation*}
  \lim_{\delta\rightarrow 0^+}\frac{\big(\lambda^*(\delta)\big)^2}{\delta}
  = \frac{1}{K_1+\epsilon},
\end{equation*}
or equivalently,
\begin{equation*}
  \lim_{\delta\rightarrow 0^+}\frac{\lambda^*(\delta)}{\sqrt{\delta}}
  = \frac{1}{\sqrt{K_1+\epsilon}}.
\end{equation*}
Thus we have
\begin{equation} \label{eq:lambdaStarDelta}
  \lambda^*(\delta)=\sqrt{\frac{\delta}{K_1+\epsilon}}+o(\sqrt{\delta}).
\end{equation}

Now consider $I_\lambda(X_1,X_2;Y)$ for $\lambda\in [0,1]$. Direct
differentiation gives \cite[Lemma 15, Part (iii)]{kUserMAC}
\begin{equation*}
  I_\lambda(X_1,X_2;Y)=I_0(X_1,X_2;Y)+
  \Big(I_1(X_1,X_2;Y)-I_0(X_1,X_2;Y)+ D\big(p_1(y)\|p_0(y)\big)\Big)\lambda
  +o(\lambda).
\end{equation*}
If we now set $\lambda=\lambda^*(\delta)$, define constant $K$ as
\begin{equation*}
  K\coloneqq \frac{1}{\sqrt{K_1+\epsilon}}
  \Big(I_1(X_1,X_2;Y)-I_0(X_1,X_2;Y)+ D\big(p_1(y)\|p_0(y)\big)\Big),
\end{equation*}
and apply (\ref{eq:lambdaStarDelta}), we get
\begin{equation*}
  I_{\lambda^*(\delta)}(X_1,X_2;Y)
  =I_0(X_1,X_2;Y)+K\sqrt{\delta}+o(\sqrt{\delta}).
\end{equation*}
Note that by the definition of $\sigma_1$,
\begin{align*}
  \sigma_1(\delta) &\geq I_{\lambda^*(\delta)}(X_1,X_2;Y)\\
  \sigma_1(0) &= I_0(X_1,X_2;Y).
\end{align*}
This concludes the proof of Lemma \ref{lem:sigmaOneLowerBound}.

\subsection{Proof of Corollary \ref{cor:discontinuity}}
\label{subsec:corDiscontinuityProof}

By Proposition \ref{prop:discontinuity}, we only need to prove 
one direction. Specifically,
here we show that for a fixed MAC and $\mathbf{C}_\mathrm{in}\in\mathbb{R}^2_{>0}$,
if $C_\mathrm{sum,max}(\mathbf{C}_\mathrm{in},
\mathbf{C}_\mathrm{out})$
is not continuous at 
$\mathbf{C}_\mathrm{out}=\mathbf{0}$, then
\begin{equation*} 
  C_\mathrm{sum}(\mathbf{0},\mathbf{0})
  >C_\mathrm{sum,max}(\mathbf{0},\mathbf{0}).
\end{equation*} 
This is equivalent to showing that if
\begin{equation}  \label{eq:avgEqualsMax}
  C_\mathrm{sum}(\mathbf{0},\mathbf{0})
  =C_\mathrm{sum,max}(\mathbf{0},\mathbf{0}),
\end{equation}
then $C_\mathrm{sum,max}(\mathbf{C}_\mathrm{in},
\mathbf{C}_\mathrm{out})$
is continuous at 
$\mathbf{C}_\mathrm{out}=\mathbf{0}$.

We begin by defining the function 
$f\colon\mathbb{R}_{\geq 0}\rightarrow
\mathbb{R}_{\geq 0}$ as 
\begin{equation*}
  f(C_\mathrm{out}) \coloneqq
  C_\mathrm{sum,max}(\mathbf{C}_\mathrm{in}^*,
  (C_\mathrm{out},C_\mathrm{out})),
\end{equation*}
where $\mathbf{C}_\mathrm{in}^*$ is defined in Section
\ref{sec:results}. 
Then by \cite[Theorem 1]{reliability} for all 
$C_\mathrm{out}>0$, and by
(\ref{eq:avgEqualsMax}) for
$C_\mathrm{out}=0$, we have
\begin{equation*}
  f(C_\mathrm{out})=
  C_\mathrm{sum}(\mathbf{C}_\mathrm{in}^*,
  (C_\mathrm{out},C_\mathrm{out})).
\end{equation*}
For each $\mathbf{C}_\mathrm{out}\in\mathbb{R}^2_{\geq 0}$,
let $C_\mathrm{out}^*\coloneqq\max\{C_\mathrm{out}^1,C_\mathrm{out}^2\}$.
Then for any $\mathbf{C}_\mathrm{in}\in\mathbb{R}^2_{\geq 0}$,
\begin{equation*}
  f(0)\leq C_\mathrm{sum,max}(\mathbf{C}_\mathrm{in},
  \mathbf{C}_\mathrm{out})\leq f(C_\mathrm{out}^*).
\end{equation*}
If we now let $\mathbf{C}_\mathrm{out}\rightarrow\mathbf{0}$
and apply Theorem \ref{thm:CsumAvgContinuity}, the continuity of 
$C_\mathrm{sum,max}(\mathbf{C}_\mathrm{in},
\mathbf{C}_\mathrm{out})$
at $\mathbf{C}_\mathrm{out}=\mathbf{0}$ follows.  

\subsection{Proof of Lemma \ref{lem:superadditivity}}
\label{subsec:superadditiveProof}

Our goal here is to show that for integers $m,n\geq 1$ and
$\delta\geq 0$,
\begin{equation*}
  (m+n)\sigma_{m+n}(\delta)\geq
  m\sigma_m(\delta)+n\sigma_n(\delta).
\end{equation*}

By the definition of $\sigma_n(\delta)$, for all $\epsilon>0$, 
there exist finite alphabets 
$\mathcal{U}_0$ and $\mathcal{U}_1$ 
and distributions $p_n\in \mathcal{P}^{(n)}_{\mathcal{U}_0}(\delta)$
and $p_m\in \mathcal{P}^{(m)}_{\mathcal{U}_1}(\delta)$ such that
\begin{align*}
  I_n(X_1^n,X_2^n;Y^n|U_0) &\geq n\sigma_n(\delta)-n\epsilon\\
  I_m(X_1^m,X_2^m;Y^m|U_1) &\geq m\sigma_m(\delta)-m\epsilon.
\end{align*}
Consider the distribution
\begin{equation*}
  p_{n+m}(u_0,u_1,x_1^{n+m},x_2^{n+m})
  =p_n(u_0,x_1^n,x_2^n)p_m(u_1,x_1^{n+1:n+m},x_2^{n+1:n+m}).
\end{equation*}
Let $\mathcal{U}\coloneqq\mathcal{U}_0\times\mathcal{U}_1$.
Then it is straightforward to show that
$p_{n+m}\in\mathcal{P}_\mathcal{U}^{(n+m)}(\delta)$, and 
\begin{equation*}
  I_{n+m}(X_1^{n+m},X_2^{n+m};Y^{n+m}|U_0,U_1)
  \geq n\sigma_n(\delta)+m\sigma_m(\delta)-(n+m)\epsilon,
\end{equation*}
which implies the desired result. 

\subsection{Proof of Lemma \ref{lem:CsumAvgBounds}}
\label{subsec:CsumAvgBounds}

Here we prove lower and upper bounds for 
$C_\mathrm{sum}(\mathbf{C}_\mathrm{in}^*,\mathbf{C}_\mathrm{out})$
in terms of $\sigma(C_\mathrm{out}^1+C_\mathrm{out}^2)$.

We first prove the lower bound. For $i\in\{1,2\}$,
choose $C_\mathrm{dep}^i$ such that 
\begin{equation*}
0\leq C_\mathrm{dep}^i\leq C_\mathrm{out}^i.
\end{equation*}
The rate $C_\mathrm{dep}^i$ indicates the amount of information
the CF sends to encoder $i$ to enable dependence between
the encoders' codewords. 
(See \cite[Section III]{kUserMAC} for details.)
For a finite alphabet $\mathcal{U}$ and positive integer $n$,
let $p(u,x_1^n,x_2^n)$ be a distribution that satisfies
\begin{equation*}
  I(X_1^n;X_2^n|U)=n(C_\mathrm{dep}^1+C_\mathrm{dep}^2).
\end{equation*}
Then applying \cite[Corollary 8]{kUserMAC}
to the MAC
\begin{equation*}
  p(y^n|x_1^n,x_2^n)=\prod_{t\in [n]}
  p(y_t|x_{1t},x_{2t}),
\end{equation*}
gives
\begin{align}
  nC_\mathrm{sum}(\mathbf{C}_\mathrm{in}^*,
  \mathbf{C}_\mathrm{out})
  &\geq I(X_1^n,X_2^n;Y^n|U)-n\min\{C_{1d},C_{2d}\}\notag\\
  &\geq I(X_1^n,X_2^n;Y^n|U)-n\min\{C_\mathrm{out}^1,C_\mathrm{out}^2\}.
  \label{eq:CsumLBCout}
\end{align}
This completes the proof of the lower bound.

For the upper bound, consider a sequence of 
$(2^{nR_1},2^{nR_2},n)$-codes for the MAC 
with a $(\mathbf{C}_\mathrm{in}^*,\mathbf{C}_\mathrm{out})$-CF.
For a fixed positive integer $n$, let $W_1$ and $W_2$
be independent random variables distributed uniformly on
$[2^{nR_1}]$ and $[2^{nR_2}]$, respectively. Furthermore, 
for the blocklength-$n$ code and transmitted message pair $(W_1,W_2)$, 
let random variables $Z_1\in [2^{nC_\mathrm{out}^1}]$ and 
$Z_2\in [2^{nC_\mathrm{out}^2}]$ denote the 
information the CF sends to encoders 1 and 2 respectively.
Note that $(Z_1,Z_2)$ is a deterministic function of $(W_1,W_2)$.
By the data processing inequality,
\begin{align*}
  I(X_1^n;X_2^n) &\leq I(X_1^n;W_2,Z_2)\\
  &\leq I(W_1,Z_1;W_2,Z_2)\\
  &= H(W_1,Z_1)+H(W_2,Z_2)-H(W_1,W_2,Z_1,Z_2)\\
  &= H(Z_1|W_1)+H(Z_2|W_2)\\
  &\leq n(C_\mathrm{out}^1+C_\mathrm{out}^2).
\end{align*}
In addition, from Fano's inequality it follows that 
there exists a sequence $(\epsilon_n)_{n=1}^\infty$
such that 
\begin{equation*}
  H(W_1,W_2|Y^n)\leq n\epsilon_n,
\end{equation*}
and $\epsilon_n\rightarrow 0$ as $n\rightarrow\infty$. 
We have
\begin{align*}
  n(R_1+R_2) &= H(W_1,W_2)\\
  &= I(W_1,W_2;Y^n)+H(W_1,W_2|Y^n)\\
  &\leq I(W_1,W_2;Y^n)+n\epsilon_n\\
  &= I(X_1^n,X_2^n;Y^n)+n\epsilon_n\\
  &\leq n\sigma(C_\mathrm{out}^1+C_\mathrm{out}^2)
  +n\epsilon_n.
\end{align*}
Dividing by $n$ and taking the limit as $n\rightarrow\infty$
completes the proof. 

\subsection{Proof of Lemma \ref{lem:concavity}}
\label{subsec:concavity}

For $n\geq 1$, we use the auxiliary random variable $U$ in the definition 
of $\sigma_n$ to show that $\sigma_n$ is concave.

It suffices to prove the result for $n=1$ since the proof for 
arbitrary integers follows similarly. 
We apply the technique from \cite[Appendix B]{CoverElGamalSalehi}.
Recall that
\begin{equation*}
  \sigma_1(\delta)= \sup_\mathcal{U}
  \max_{p\in \mathcal{P}^{(1)}_\mathcal{U}(\delta)}
  I(X_1,X_2;Y|U). 
\end{equation*}
Fix $a,b\geq 0$, $\lambda\in (0,1)$, and $\epsilon>0$. 
By definition, there exist finite sets $\mathcal{U}_0$ and $\mathcal{U}_1$
and distributions
$p_0\in \mathcal{P}^{(1)}_{\mathcal{U}_0}(a)$
and $p_1\in \mathcal{P}^{(1)}_{\mathcal{U}_1}(b)$ 
satisfying 
\begin{align*}
  I_0(X_1,X_2;Y|U_0) &\geq \sigma_1(a)-\epsilon\\
  I_1(X_1,X_2;Y|U_1) &\geq \sigma_1(b)-\epsilon,
\end{align*}
respectively. Define the alphabet $\mathcal{V}$ as
\begin{equation*}
  \mathcal{V}\coloneqq \{0\}\times\mathcal{U}_0
  \cup \{1\}\times\mathcal{U}_1.
\end{equation*}
We denote an element of $\mathcal{V}$ by
$v=(v_1,v_2)$. 
Define the distribution $p_\lambda(v,x_1,x_2)$ as
\begin{equation*}
  p_\lambda(v,x_1,x_2)
  =p_\lambda(v_1)
  p_{v_1}(v_2,x_1,x_2),
\end{equation*}
where 
\begin{equation*}
  p_\lambda(v_1)
  =\begin{cases}
  1-\lambda &\text{if }v_1=0\\
  \lambda &\text{if }v_1=1.
  \end{cases}
\end{equation*}
Then
\begin{align*}
  I_\lambda(X_1;X_2|V)
  &= I_\lambda(X_1,X_2|V_1,V_2)\\
  &= (1-\lambda)I(X_1;X_2|V_1=0,V_2)
  +\lambda I(X_1;X_2|V_1=1,V_2)\\
  &= (1-\lambda)I_0(X_1;X_2|U_0)
  +\lambda I_1(X_1;X_2|U_1)\\
  &\leq (1-\lambda)a+\lambda b,
\end{align*}
which implies 
$p_\lambda\in\mathcal{P}^{(1)}_\mathcal{V}
((1-\lambda)a+\lambda b)$. Similarly,
\begin{align*}
  I_\lambda(X_1,X_2;Y|V)
  &= I_\lambda(X_1,X_2;Y|V_1,V_2)\\
  &= (1-\lambda)I(X_1,X_2;Y|V_1=0,V_2)
  +\lambda I(X_1,X_2;Y|V_1=1,V_2)\\
  &= (1-\lambda)I_0(X_1,X_2;Y|U_0)
  +\lambda I_1(X_1,X_2;Y|U_1)\\
  &\geq (1-\lambda)\sigma_1(a)+\lambda \sigma_1(b)-\epsilon.
\end{align*}
Therefore, 
\begin{equation*}
  \sigma_1\big((1-\lambda)a+\lambda b\big)\geq 
  (1-\lambda)\sigma_1(a)+\lambda \sigma_1(b)-\epsilon.
\end{equation*}
The result now follows from the fact that the above
equation holds for all $\epsilon >0$. 

\subsection{Proof of Lemma \ref{lem:Dueck} (Dueck's Lemma)} 
\label{subsec:Dueck}

If for all $t\in [n]$, we have
\begin{equation*}
 I(X_{1t};X_{2t}|U)\leq \epsilon,
\end{equation*}
then we define $T\coloneqq \varnothing$. Otherwise,
there exists $t_1\in [n]$ such that
\begin{equation} \label{eq:singleLetterBoundMI}
 I(X_{1t_1};X_{2t_1}|U)>\epsilon.
\end{equation}
Let $S_1\coloneqq [n]\setminus\{t_1\}$. Then 
\begin{align*}
 I(X_1^n;X_2^n|U) &= I(X_1^n;X_{2t_1}|U)+I(X_1^n;X_2^{S_1}|U,X_{2t_1})\\
 &= I(X_{1t_1};X_{2t_1}|U)+I(X_1^{S_1};X_{2t_1}|U,X_{1t_1})\\
 &\phantom{=}+I(X_{1t_1};X_{2}^{S_1}|U,X_{2t_1})+I(X_1^{S_1};X_2^{S_1}|U,X_{1t_1},X_{2t_1})\\
 &\geq I(X_{1t_1};X_{2t_1}|U)+I(X_1^{S_1};X_2^{S_1}|U,X_{1t_1},X_{2t_1}).
\end{align*}
Since $I(X_1^n;X_2^n|U)\leq n\delta$, using (\ref{eq:singleLetterBoundMI}),
we get
\begin{equation*}
 I(X_1^{S_1};X_2^{S_1}|U,X_{1t_1},X_{2t_1})
 \leq n\delta-\epsilon.
\end{equation*}
Now if for all $t\in S_1$, 
\begin{equation*}
 I(X_{1t};X_{2t}|U,X_{1t_1},X_{2t_1})\leq\epsilon,
\end{equation*}
then we define $T\coloneqq \{t_1\}$. Otherwise, there exists $t_2\in [n]$ such that 
\begin{equation*}
 I(X_{1t_2};X_{2t_2}|U,X_{1t_1},X_{2t_1})>\epsilon.
\end{equation*}
Similar to the above argument, if we define
$S_2\coloneqq [n]\setminus\{t_1,t_2\}$,  then
\begin{equation*} I(X_1^{S_2};X_2^{S_2}|U,X_{1t_1},X_{1t_2},X_{2t_1},X_{2t_2})
 \leq n\delta-2\epsilon.
\end{equation*}
If we continue this process, we eventually get a set  $T\coloneqq \{t_1,\dots,t_k\}$ such that
\begin{equation} \label{eq:SkTeps}
 I(X_1^{T^c};X_2^{T^c}|U,X_{1}^T,X_{2}^T)
 \leq n\delta-|T|\epsilon,
\end{equation}
and for all $t\in S_k\coloneqq T^c$,
\begin{equation*}
 I(X_{1t};X_{2t}|U,X_{1}^T,X_{2}^T)\leq\epsilon.
\end{equation*}
In addition, from (\ref{eq:SkTeps}) it follows that
\begin{equation}
 |T|\leq \frac{n\delta}{\epsilon}.
\end{equation}

\subsection{Proof of Corollary \ref{cor:DueckWringing}} 
\label{subsec:CorDueck}

Here we give an upper bound for $\sigma(\delta)$, which is defined as 
\begin{equation*}
 \sigma(\delta)\coloneqq
 \lim_{n\rightarrow\infty}
 \sigma_n(\delta).
\end{equation*}
in terms of $\sigma_1(\delta)$.

Fix a positive integer $n$. By Lemma \ref{lem:cardinality}, we
can choose $\mathcal{U}=\{a,b\}$ in the definition of 
$\sigma_n(\delta)$. 
From Lemma \ref{lem:Dueck}, it follows that there exists
a set $T\subseteq [n]$ such that
\begin{equation} \label{eq:boundCardinalityT}
 0\leq |T|\leq \frac{n\delta}{\epsilon},
\end{equation}
and 
\begin{equation*}
 \forall\: t\notin T\colon
 I(X_{1t};X_{2t}|U,X_1^T,X_2^T)\leq\epsilon.
\end{equation*}
Thus
\begin{align}
 I(X_1^n,X_2^n;Y^n|U)
 &= I(X_1^T,X_2^T;Y^n|U)+I(X_1^{T^c},X_2^{T^c};Y^n|U,X_1^T,X_2^T)
 \notag\\
 &\leq |T|\log |\mathcal{X}_1||\mathcal{X}_2|
 +I(X_1^{T^c},X_2^{T^c};Y^n|U,X_1^T,X_2^T).
 \label{eq:boundInputOutputMI}
\end{align}
We further bound the second term on the right hand side by 
\begin{align}
 \MoveEqLeft
 I(X_1^{T^c},X_2^{T^c};Y^n|U,X_1^T,X_2^T)\notag\\
 &= I(X_1^{T^c},X_2^{T^c};Y^{T^c}|U,X_1^T,X_2^T)
 +I(X_1^{T^c},X_2^{T^c};Y^T|U,X_1^T,X_2^T,Y^{T^c})\notag\\
 &= I(X_1^{T^c},X_2^{T^c};Y^{T^c}|U,X_1^T,X_2^T)\notag\\
 &\leq \sum_{t\notin T}
 I(X_{1t},X_{2t};Y_t|U,X_1^T,X_2^T)\notag\\
 &\leq n\max_{p\in\mathcal{P}^{(1)}_{\mathcal{V}}(\epsilon)}
 I(X_1,X_2;Y|V)\leq n\sigma_1(\epsilon),
 \label{eq:TcYBoundConditionedT}
\end{align}
where in (\ref{eq:TcYBoundConditionedT}), 
\begin{equation*}
 \mathcal{V}\coloneqq \mathcal{U}\times\mathcal{X}_1^{|T|}\times\mathcal{X}_2^{|T|}.
\end{equation*}
Therefore, by (\ref{eq:boundCardinalityT}), (\ref{eq:boundInputOutputMI}), 
and (\ref{eq:TcYBoundConditionedT}),
\begin{equation*}
 \frac{1}{n}I(X_1^n,X_2^n;Y^n|U)\leq
 \frac{\delta}{\epsilon}\log 
 |\mathcal{X}_1||\mathcal{X}_2|
 +\sigma_1(\epsilon),
\end{equation*}
which completes the proof.

\subsection{Proof of Lemma \ref{lem:cardinality}}
\label{subsec:cardinalityProof}

Here we show that in the definition of $\sigma_1$, 
\begin{equation*}
 \sigma_1(\delta)\coloneqq
 \sup_{\mathcal{U}}
 \max_{p\in\mathcal{P}^{(n)}_\mathcal{U}(\delta)}
 I(X_1,X_2;Y|U)
\end{equation*}
where the supremum is over all finite sets $\mathcal{U}$, 
we can instead take the supremum over all sets $\mathcal{U}$ with 
cardinality at most two. 

Let $\mathcal{U}$ be some 
finite set and let $p^*(u,x_1,x_2)\in\mathcal{P}^{(1)}_\mathcal{U}(\delta)$
be a distribution that satisfies
\begin{equation*}
 I^*(X_1,X_2;Y|U)=\max_{p\in \mathcal{P}^{(1)}_\mathcal{U}(\delta)}
 I(X_1,X_2;Y|U),
\end{equation*}
where the mutual information term $I^*(X_1,X_2;Y|U)$ on the
left hand side is calculated with respect to 
$p^*(u,x_1,x_2)p(y|x_1,x_2)$.
Let $\mathcal{Q}\subseteq\mathbb{R}^{|\mathcal{U}|}$ denote the set of all
vectors $(q(u))_{u\in\mathcal{U}}$ that satisfy
\begin{align}
 &q(u)\geq 0\text{ for all }u\in\mathcal{U}\notag\\
 &\sum_{u\in\mathcal{U}}q(u)=1\notag\\
 &\sum_{u\in\mathcal{U}}q(u)I^*(X_1;X_2|U=u)
 =I^*(X_1;X_2|U),\label{eq:qMutualInfo}
\end{align}
where in (\ref{eq:qMutualInfo}), $I^*(X_1;X_2|U=u)$
and $I^*(X_1;X_2|U)$ are calculated according to
$p^*(x_1,x_2|u)$ and $p^*(u,x_1,x_2)$, respectively. 
Note that $\mathcal{Q}$ is nonempty since $p^*(u)\in\mathcal{Q}$.
Consider the mapping $F\colon\mathcal{Q}\rightarrow\mathbb{R}_{\geq 0}$ 
defined by
\begin{equation} \label{eq:mappingFdef}
 F[q]\coloneqq 
 \sum_{u\in\mathcal{U}}q(u)I^*(X_1,X_2;Y|U=u),
\end{equation}
where $I^*(X_1,X_2;Y|U=u)$ is calculated with respect to 
$p^*(x_1,x_2|u)p(y|x_1,x_2)$. Since
$p^*(u)\in\mathcal{Q}$ and for all $q(u)\in\mathcal{Q}$, by 
(\ref{eq:qMutualInfo}) we have
$q(u)p^*(x_1,x_2|u)\in\mathcal{P}^{(1)}_\mathcal{U}(\delta)$,
thus
\begin{equation*}
 \max_{q\in\mathcal{Q}}F[q]= 
 \max_{p\in \mathcal{P}^{(1)}_\mathcal{U}(\delta)}
 I(X_1,X_2;Y|U).
\end{equation*}
Therefore, it suffices to find $q^*\in\mathcal{Q}$ which has at most
two non-zero components and at which $F$ 
obtains its maximal value. To this end, note that since
$\mathcal{Q}$ is a nonempty bounded polyhedron, by
\cite[p. 65, Corollary 2.2]{BertsimasTsitsiklis} 
and \cite[p. 50, Theorem 2.3]{BertsimasTsitsiklis},
$\mathcal{Q}$ has at least one extreme point. 
Since $F$ is linear in $q$ and $\mathcal{Q}$ has at least one extreme point,
\cite[p. 65, Theorem 2.7]{BertsimasTsitsiklis} shows that there 
exists an extreme point of
$\mathcal{Q}$, say $q^*\in\mathcal{Q}$, at which $F$ obtains its
maximum. Finally, since $q^*$ is an extreme point, applying
\cite[p. 50, Theorem 2.3]{BertsimasTsitsiklis} to the definition
of $\mathcal{Q}$ implies that
we must have $q^*(u)=0$ for at least 
$|\mathcal{U}|-2$ values of $u$. This completes the proof. 

\subsection{Proof of Lemma \ref{lem:continuityFn}}
\label{subsec:continuityFnProof}

We next show that $\sigma_1(\delta)$ is continuous at $\delta=0$.

By Lemma \ref{lem:cardinality}, without loss of generality, we can
set $\mathcal{U}\coloneqq\{a,b\}$. (Our proof below applies for any
finite set $\mathcal{U}$.)
For all $\delta\geq 0$, we have
\begin{equation*}
 \sigma_1(\delta)= 
 \max_{p\in \mathcal{P}^{(1)}_\mathcal{U}(\delta)}
 I(X_1,X_2;Y|U). 
\end{equation*}
Fix $\delta>0$. Let $p^*(u,x_1,x_2)$ be a distribution 
in $\mathcal{P}_\mathcal{U}^{(1)}(\delta)$ achieving
the maximum above, and define
\begin{equation*}
 p^*_\mathrm{ind}(x_1,x_2|u)\coloneqq 
 p^*(x_1|u)p^*(x_2|u).
\end{equation*}
In addition, for each $u\in\mathcal{U}$, let
\begin{align*}
 D\big(p^*(x_1,x_2|u)\|p^*_\mathrm{ind}(x_1,x_2|u)\big)
 &= \sum_{x_1,x_2}p^*(x_1,x_2|u)
 \log\frac{p^*(x_1,x_2|u)}{p^*_\mathrm{ind}(x_1,x_2|u)}\\
 \big\|p^*(y|u)-p^*_\mathrm{ind}(y|u)\big\|_{L^1}
 &= \sum_y \big|p^*(y|u)-p^*_\mathrm{ind}(y|u)\big|\\
 \big\| p^*(x_1,x_2|u)-p^*_\mathrm{ind}(x_1,x_2|u)\big\|_{L^1}
 &= \sum_{x_1,x_2}\big|p^*(x_1,x_2|u)-p^*_\mathrm{ind}(x_1,x_2|u)\big|.
\end{align*}

Since
\begin{equation*}
 \sum_{u\in\mathcal{U}}
 p^*(u)D\big(p^*(x_1,x_2|u)\|p^*_\mathrm{ind}(x_1,x_2|u)\big)
 = I^*(X_1;X_2|U)\leq \delta,
\end{equation*}
by applying \cite[Lemma 11.6.1]{CoverThomas} for every $u\in\mathcal{U}$,
we get
\begin{equation} \label{eq:L1bound}
 \sum_{u\in\mathcal{U}}p^*(u)
 \big\| p^*(x_1,x_2|u)-p^*_\mathrm{ind}(x_1,x_2|u)\big\|^2_{L^1}
 \leq 2\delta\ln 2.
\end{equation}
In addition,  
\begin{align}
 \MoveEqLeft
 \sum_{u\in\mathcal{U}}p^*(u)
 \big\|p^*(y|u)-p^*_\mathrm{ind}(y|u)\big\|_{L^1}\notag\\
 &\leq \sum_{u\in\mathcal{U}}p^*(u)\sum_{x_1,x_2}p(y|x_1,x_2)\big|
 p^*(x_1,x_2|u)-p^*_\mathrm{ind}(x_1,x_2|u)\big|\notag\\
 &\leq \sum_{u\in\mathcal{U}}p^*(u)
 \big\| p^*(x_1,x_2|u)-p^*_\mathrm{ind}(x_1,x_2|u)\big\|_{L^1}
 \notag\\
 &\leq \sqrt{2\delta\ln 2},\label{eq:twoDelta}
\end{align}
where (\ref{eq:twoDelta}) follows from (\ref{eq:L1bound}) and
the Cauchy-Schwarz inequality. 
Define the subset $\mathcal{U}_0\subseteq\mathcal{U}$ as
\begin{equation*}
 \mathcal{U}_0=\Big\{u\in\mathcal{U}:
 \big\|p^*(y|u)-p^*_\mathrm{ind}(y|u)\big\|_{L^1}
 \leq 1/2\Big\}.
\end{equation*}
Clearly, by (\ref{eq:twoDelta}),
\begin{equation} \label{eq:PrNotInU0}
 \sum_{u\notin\mathcal{U}_0}p^*(u)
 \leq 2\sqrt{2\delta\ln 2}.
\end{equation}
Thus
\begin{align}
 \MoveEqLeft
 \big|H^*(Y|U)-H^*_\mathrm{ind}(Y|U)\big|\notag\\
 &\leq \sum_{u\in\mathcal{U}}p^*(u)
 \big|H^*(Y|U=u)-H^*_\mathrm{ind}(Y|U=u)\big|\notag\\
 &\leq 2\sqrt{2\delta\ln 2}\log |\mathcal{Y}|
 \label{eq:entropy2L1}\\
 &\phantom{\leq}
 -\sum_{u\in\mathcal{U}_0}p^*(u)
 \big\|p^*(y|u)-p^*_\mathrm{ind}(y|u)\big\|_{L^1}
 \log\frac{\big\|p^*(y|u)-p^*_\mathrm{ind}(y|u)\big\|_{L^1}}
 {|\mathcal{Y}|}\notag\\
 &\leq 2\sqrt{2\delta\ln 2}\log |\mathcal{Y}|
 \label{eq:U0toFullSupport}\\
 &\phantom{\leq}
 -\sum_{u\in\mathcal{U}}p^*(u)
 \big\|p^*(y|u)-p^*_\mathrm{ind}(y|u)\big\|_{L^1}
 \log\frac{\big\|p^*(y|u)-p^*_\mathrm{ind}(y|u)\big\|_{L^1}}
 {|\mathcal{Y}|}\notag\\
 &\leq 2\sqrt{2\delta\ln 2}\log |\mathcal{Y}|-\sqrt{2\delta\ln 2}\log\Big(
 \frac{1}{|\mathcal{Y}|}\sqrt{2\delta\ln 2}\Big)\label{eq:smallDeltaBound}\\
 &=\sqrt{2\delta\ln 2}\log\frac{|\mathcal{Y}|^3}{\sqrt{2\delta\ln 2}},\label{eq:EntropyBoundDifference}
\end{align}
where $(\ref{eq:entropy2L1})$ follows from (\ref{eq:PrNotInU0}) and 
\cite[Theorem 17.3.3]{CoverThomas}, $(\ref{eq:U0toFullSupport})$ follows from the fact
that for all $u\in\mathcal{U}$,
\begin{equation*}
 -\big\|p^*(y|u)-p^*_\mathrm{ind}(y|u)\big\|_{L^1}
 \log\frac{\big\|p^*(y|u)-p^*_\mathrm{ind}(y|u)\big\|_{L^1}}
 {|\mathcal{Y}|}\geq 0,
\end{equation*}
and for $\delta$ satisfying
\begin{equation} \label{eq:smallDeltaConstraint}
0<\sqrt{2\delta\ln 2}<\frac{1}{e}|\mathcal{Y}|,
\end{equation}
$(\ref{eq:smallDeltaBound})$ follows from (\ref{eq:twoDelta}) and the fact that the mapping 
$t\mapsto -t\log (t/|\mathcal{Y}|)$ is concave on $(0,\infty)$ and increasing on 
$(0,|\mathcal{Y}|/e)$. In addition, by (\ref{eq:twoDelta}),
\begin{align}
 \MoveEqLeft
 \big|H^*(Y|U,X_1,X_2)-H^*_\mathrm{ind}(Y|U,X_1,X_2)\big|\notag\\
 &\leq \sum_{u,x_1,x_2}
 \big|p^*(u,x_1,x_2)-p^*_\mathrm{ind}(u,x_1,x_2)\big|H(Y|X_1=x_1,X_2=x_2)
 \notag\\
 &\leq \sum_{u,x_1,x_2}
 \big|p^*(u,x_1,x_2)-p^*_\mathrm{ind}(u,x_1,x_2)\big|\log|\mathcal{Y}|
 \notag\\
 &= \Big(\log|\mathcal{Y}|\Big)\cdot
 \sum_{u\in\mathcal{U}}p^*(u)
 \big\|p^*(x_1,x_2|u)-p^*_\mathrm{ind}(x_1,x_2|u)\big\|_{L^1}
 \notag\\
 &\leq \Big(\log|\mathcal{Y}|\Big)\sqrt{2\delta\ln 2}.
 \label{eq:logYTwoDelta}
\end{align}
Thus by (\ref{eq:EntropyBoundDifference}) and
(\ref{eq:logYTwoDelta}), for all $\delta$ satisfying
(\ref{eq:smallDeltaConstraint}), 
\begin{align*}
 \sigma_1(\delta) &= I^*(X_1,X_2;Y|U)
 =H^*(Y|U)-H^*(Y|U,X_1,X_2)\\
 &\leq \big|H^*(Y|U)-H^*_\mathrm{ind}(Y|U)\big|
 +\big|H^*(Y|U,X_1,X_2)-H^*_\mathrm{ind}(Y|U,X_1,X_2)\big|\\
 &\phantom{=}
 +I^*_\mathrm{ind}(X_1,X_2;Y|U)\\
 &\leq \sqrt{2\delta\ln 2}\log\frac{|\mathcal{Y}|^3}{\sqrt{2\delta\ln 2}}
 +\Big(\log|\mathcal{Y}|\Big)\sqrt{2\delta\ln 2}
 +\sigma_1(0).
\end{align*}
Since $\sigma_1(0)\leq \sigma_1(\delta)$ for all $\delta\geq 0$, 
the continuity of $\sigma_1$ at $\delta=0^+$ follows. 

\subsection{Proof of Lemma \ref{lem:continuityCin}}
\label{subsec:continuityCin}
For a fixed $\mathbf{C}_\mathrm{out}\in\mathbb{R}^2_{\geq 0}$,
here we show that 
$C_\mathrm{sum}(\mathbf{C}_\mathrm{in}, \mathbf{C}_\mathrm{out})$
is continuous with respect to $\mathbf{C}_\mathrm{in}$. 

Define the functions $f,g\colon\mathbb{R}^2_{\geq 0}\rightarrow\mathbb{R}$
as 
\begin{align*}
 f(\mathbf{C}_\mathrm{in}) &\coloneqq C_\mathrm{sum}(\mathbf{C}_\mathrm{in},\mathbf{C}_\mathrm{out})
 -C_\mathrm{sum}(\mathbf{C}_\mathrm{in},\mathbf{0})
 =C_\mathrm{sum}(\mathbf{C}_\mathrm{in},\mathbf{C}_\mathrm{out})
 -C_\mathrm{sum}(\mathbf{0},\mathbf{0})\\
 g(\mathbf{C}_\mathrm{in}) &\coloneqq
 C_\mathrm{in}^1+C_\mathrm{in}^2-f(\mathbf{C}_\mathrm{in}).
\end{align*}
Note that since $f$ is concave, $g$ is convex. Thus 
for all $\lambda\in [0,1]$ and all 
$(\mathbf{C}_\mathrm{in},\mathbf{\tilde{C}}_\mathrm{in})$,
\begin{equation*}
 g(\lambda\mathbf{C}_\mathrm{in}
 +(1-\lambda)\mathbf{\tilde{C}}_\mathrm{in})
 \leq \lambda g(\mathbf{C}_\mathrm{in})
 +(1-\lambda)g(\mathbf{\tilde{C}}_\mathrm{in}).
\end{equation*}
Since $g(\mathbf{0})=0$, setting 
$\mathbf{\tilde{C}}_\mathrm{in}=\mathbf{0}$
gives 
\begin{equation*}
 g(\lambda\mathbf{C}_\mathrm{in})
 \leq \lambda g(\mathbf{C}_\mathrm{in})
\end{equation*}
Note that by \cite[Proposition 6]{kUserMAC}, $g$ is nonnegative. Thus
\begin{equation*}
 g(\lambda\mathbf{C}_\mathrm{in})
 \leq g(\mathbf{C}_\mathrm{in}),
\end{equation*}
which when written in terms of $f$, is equivalent
to
\begin{equation} \label{eq:sumCapacityBoundCin}
 f(\mathbf{C}_\mathrm{in})
 -f(\lambda\mathbf{C}_\mathrm{in})
 \leq (1-\lambda)(C_\mathrm{in}^1+C_\mathrm{in}^2).
\end{equation}

Consider 
$\mathbf{C}_\mathrm{in},\mathbf{\tilde{C}}_\mathrm{in}
\in\mathbb{R}^2_{\geq 0}$. Define the pairs 
$\mathbf{\underaccent{\bar} C}_\mathrm{in},\mathbf{\bar C}_\mathrm{in}
\in\mathbb{R}^2_{\geq 0}$ as
\begin{align*}
 &\forall\:i\in\{1,2\}\colon
 \underaccent{\bar}{C}_\mathrm{in}^i
 \coloneqq \min\{C_\mathrm{in}^i,\tilde{C}_\mathrm{in}^i\}\\
 &\forall\:i\in\{1,2\}\colon
 \bar{C}_\mathrm{in}^i
 \coloneqq \max\{C_\mathrm{in}^i,\tilde{C}_\mathrm{in}^i\}.\\
\end{align*}
Next define 
$\lambda^*(\mathbf{C}_\mathrm{in},\mathbf{\tilde{C}}_\mathrm{in})\in [0,1]$ 
as\footnote{If for some $i\in\{1,2\}$, say $i=1$, 
$\bar{C}_\mathrm{in}^i=0$, set $\lambda^*\coloneqq
\min\{1,\underaccent{\bar}{C}_\mathrm{in}^2/\bar{C}_\mathrm{in}^2\}$.
If $\bar{C}_\mathrm{in}^1=\bar{C}_\mathrm{in}^2=0$, set 
$\lambda^*=1$. These definitions ensure the continuity of 
$\lambda^*$ in $(\mathbf{C}_\mathrm{in},\mathbf{\tilde{C}}_\mathrm{in})$.}
\begin{equation*}
 \lambda^*\coloneqq
 \min_{i\in\{1,2\}}\underaccent{\bar}{C}_\mathrm{in}^i/
 \bar{C}_\mathrm{in}^i.
\end{equation*}
Then
\begin{align}
 \big|f(\mathbf{C}_\mathrm{in})-f(\mathbf{\tilde C}_\mathrm{in})\big|
 &\leq \big|f(\mathbf{C}_\mathrm{in})-f(\mathbf{\underaccent{\bar} C}_\mathrm{in})\big|
 +\big|f(\mathbf{\underaccent{\bar} C}_\mathrm{in})-f(\mathbf{\tilde C}_\mathrm{in})\big|
 \label{eq:continuityCin1}\\
 &=f(\mathbf{C}_\mathrm{in})-f(\mathbf{\underaccent{\bar} C}_\mathrm{in})
 +f(\mathbf{\tilde C}_\mathrm{in})-f(\mathbf{\underaccent{\bar} C}_\mathrm{in})
 \label{eq:continuityCin2}\\
 &\leq f(\mathbf{C}_\mathrm{in})-f(\lambda^*\mathbf{C}_\mathrm{in})
 +f(\mathbf{\tilde C}_\mathrm{in})-f(\lambda^*\mathbf{\tilde C}_\mathrm{in})
 \label{eq:continuityCin3}\\
 &\leq (1-\lambda^*)(C_\mathrm{in}^1+C_\mathrm{in}^2)
 +(1-\lambda^*)(\tilde{C}_\mathrm{in}^1+\tilde{C}_\mathrm{in}^2),
 \label{eq:continuityCin4}
\end{align}
where (\ref{eq:continuityCin1}) follows from the triangle inequality,
(\ref{eq:continuityCin2}) follows from the definition of 
$\mathbf{\underaccent{\bar} C}_\mathrm{in}$, 
(\ref{eq:continuityCin3}) follows from the definition of 
$\lambda^*$, and (\ref{eq:continuityCin4}) follows from
(\ref{eq:sumCapacityBoundCin}). Finally, if we let 
$\mathbf{\tilde{C}}_\mathrm{in}\rightarrow\mathbf{C}_\mathrm{in}$
in (\ref{eq:continuityCin4}), we see that 
$f(\mathbf{\tilde{C}}_\mathrm{in})\rightarrow
f(\mathbf{C}_\mathrm{in})$, since 
\begin{equation*}
 \lim_{\mathbf{\tilde{C}}_\mathrm{in}\rightarrow\mathbf{C}_\mathrm{in}}
 \lambda^*(\mathbf{C}_\mathrm{in},\mathbf{\tilde{C}}_\mathrm{in})=1.
\end{equation*}

\subsection{Proof of Lemma \ref{lem:continuityCout}}
\label{subsec:continuityCout}

Here we derive a necessary and sufficient condition for the 
continuity of $C_\mathrm{sum}(\mathbf{C}_\mathrm{in},\mathbf{C}_\mathrm{out})$
with respect to $\mathbf{C}_\mathrm{out}$ for a fixed
$\mathbf{C}_\mathrm{in}$. 

Recall that we only need to verify continuity on the boundary
of $\mathbb{R}^2_{\geq 0}\times\mathbb{R}^2_{\geq 0}$; namely,
the set of all points 
$(\mathbf{C}_\mathrm{in},\mathbf{C}_\mathrm{out})$ where at least
one of $C_\mathrm{in}^1$, $C_\mathrm{in}^2$, $C_\mathrm{out}^1$,
or $C_\mathrm{out}^2$ is zero. 

We first show that 
\begin{equation*}
 \lim_{\mathbf{\tilde{C}}_\mathrm{out}\rightarrow\mathbf{C}_\mathrm{out}} 
 C_\mathrm{sum}(\mathbf{C}_\mathrm{in},\mathbf{\tilde{C}}_\mathrm{out})
 =C_\mathrm{sum}(\mathbf{C}_\mathrm{in},\mathbf{C}_\mathrm{out}).
\end{equation*}
holds if $\mathbf{C}_\mathrm{out}=\mathbf{0}$, or if either
$C_\mathrm{in}^1=0$ or $C_\mathrm{in}^2=0$. For the case
$\mathbf{C}_\mathrm{out}=\mathbf{0}$, note that 
\begin{align*}
 C_\mathrm{sum}(\mathbf{C}_\mathrm{in},\mathbf{\tilde{C}}_\mathrm{out})
 -C_\mathrm{sum}(\mathbf{C}_\mathrm{in},\mathbf{0})
 &= C_\mathrm{sum}(\mathbf{C}_\mathrm{in},\mathbf{\tilde{C}}_\mathrm{out})
 -C_\mathrm{sum}(\mathbf{0},\mathbf{0})\\
 &\leq C_\mathrm{sum}(\mathbf{C}_\mathrm{in}^*,\mathbf{\tilde{C}}_\mathrm{out})
 -C_\mathrm{sum}(\mathbf{0},\mathbf{0}),
\end{align*}
which goes to zero as $\mathbf{\tilde{C}}_\mathrm{out}\rightarrow\mathbf{0}$
by Theorem \ref{thm:CsumAvgContinuity}.

Next suppose $C_\mathrm{in}^2=0$. In this case, we have
\begin{equation*}
 C_\mathrm{sum}\big((C_\mathrm{in}^1,0),\mathbf{\tilde{C}}_\mathrm{out}\big)
 = C_\mathrm{sum}\big((C_\mathrm{in}^1,0),(0,\tilde{C}_\mathrm{out}^2)\big).
\end{equation*}
Let $f\colon\mathbb{R}_{\geq 0}\rightarrow\mathbb{R}_{\geq 0}$
denote the function
\begin{equation*}
 f(C_\mathrm{out}^2)
 \coloneqq 
 C_\mathrm{sum}\big((C_\mathrm{in}^1,0),(0,C_\mathrm{out}^2)\big).
\end{equation*} 
Note that $f$ is continuous on $\mathbb{R}_{>0}$ since it is concave.
To prove the continuity of $f$ at $C_\mathrm{out}^2=0$, observe that
$f(C_\mathrm{out}^2)$ equals the sum-capacity of a MAC with 
a $(C_{12},0)$-conference \cite{WillemsMAC}, where 
\begin{equation*}
 C_{12}\coloneqq\min\{C_\mathrm{in}^1,C_\mathrm{out}^2\}.
\end{equation*}
From the capacity region given in \cite{WillemsMAC}, we have
\begin{equation*}
 f(C_\mathrm{out}^2)
 \leq f(0)+\min\{C_\mathrm{in}^1,C_\mathrm{out}^2\},
\end{equation*}
which implies that $f$ is 
continuous at $C_\mathrm{out}^2=0$.
The case where $C_\mathrm{in}^1=0$ follows similarly. 

Finally, consider the case where $C_\mathrm{out}^2=0$, but 
$C_\mathrm{out}^1>0$. In this case, we apply
the next lemma for concave functions that are nondecreasing as
well. 
\begin{lemma} \label{lem:concaveFunctions}
Let $f\colon\mathbb{R}_{\geq 0}\rightarrow\mathbb{R}_{\geq 0}$ be
concave and nondecreasing. Then if $|x-y|\leq \min\{x,y\}$,
\begin{equation*}
 \big|f(x)-f(y)\big|
 \leq f(|x-y|)-f(0).
\end{equation*}
\end{lemma}

We have 
\begin{align*}
 \MoveEqLeft
 \big|C_\mathrm{sum}(\mathbf{C}_\mathrm{in},(\tilde{C}_\mathrm{out}^1,\tilde{C}_\mathrm{out}^2))
 -C_\mathrm{sum}(\mathbf{C}_\mathrm{in},(C_\mathrm{out}^1,0))\big|\\
 &\overset{(a)}{\leq}
 \big|C_\mathrm{sum}(\mathbf{C}_\mathrm{in},(\tilde{C}_\mathrm{out}^1,\tilde{C}_\mathrm{out}^2))
 -C_\mathrm{sum}(\mathbf{C}_\mathrm{in},(C_\mathrm{out}^1,\tilde{C}_\mathrm{out}^2))\big|
 +\big|C_\mathrm{sum}(\mathbf{C}_\mathrm{in},(C_\mathrm{out}^1,\tilde{C}_\mathrm{out}^2))
 -C_\mathrm{sum}(\mathbf{C}_\mathrm{in},(C_\mathrm{out}^1,0))\big|\\
 &\overset{(b)}{\leq}
 \big|C_\mathrm{sum}(\mathbf{C}_\mathrm{in},(|\tilde{C}_\mathrm{out}^1-C_\mathrm{out}^1|,\tilde{C}_\mathrm{out}^2))
 -C_\mathrm{sum}(\mathbf{C}_\mathrm{in},(0,\tilde{C}_\mathrm{out}^2))\big|\\
 &\phantom{\leq}
 +\big|C_\mathrm{sum}(\mathbf{C}_\mathrm{in},(C_\mathrm{out}^1,\tilde{C}_\mathrm{out}^2))
 -C_\mathrm{sum}(\mathbf{C}_\mathrm{in},(C_\mathrm{out}^1,0))\big|,
\end{align*}
where (a) follows from the triangle inequality, and (b) follows from Lemma
\ref{lem:concaveFunctions}. If we now let 
$\mathbf{\tilde{C}}_\mathrm{out}\rightarrow (C_\mathrm{out}^1,0)$, 
Corollary \ref{cor:continuityAtZero} implies
\begin{equation*}
 \lim_{\mathbf{\tilde{C}}_\mathrm{out}\rightarrow (C_\mathrm{out}^1,0)}
 \big|C_\mathrm{sum}(\mathbf{C}_\mathrm{in},(|\tilde{C}_\mathrm{out}^1-C_\mathrm{out}^1|,\tilde{C}_\mathrm{out}^2))
 -C_\mathrm{sum}(\mathbf{C}_\mathrm{in},(0,\tilde{C}_\mathrm{out}^2))\big|=0,
\end{equation*}
from which our result follows. An analogous proof applies
in the case where $C_\mathrm{out}^1=0$ and
$C_\mathrm{out}^2>0$.

\mikel{

\subsection{Proof of Claim \ref{claim:good}}
\label{sec:claim:good}

The result follows from the definition of $W^{n_1}$.
Consider entry $i$ of $w_2\in\{0,1\}^{n_1}$, $w_{2;i}\in\{0,1\}$, and its corresponding entry $\bar{w}_{2:i}\in \{0,1\}$ in $\bar{w}_2 \in \{0,1\}^{n_1}$.
Let $\bxf_1 =f_{1;1}(w_1)$, and let $x_{1;i}$ be the $i$'th entry of $\bxf_1$.
It holds (by a simple exhaustive case analysis) that exactly one of the values $W_1(x_{1;i},w_{2;i})$ and $W_1(x_{1:i},\bar{w}_{2;i})$ is in the set $\{c,C\}$. Thus, the number of entries in $W_1^{n_1}(f_{1;1}(w_1),w_2)$ that are in the set $\{c,C\}$ plus the number of entries in $W_1^{n_1}(f_{1;1}(w_1),\bar{w}_2)$ that are in the set $\{c,C\}$ equals $n_1$. If the former exceeds $n_1/2$ (i.e., $W^{n_1}(f_{1;1}(w_1),w_2)$ is bad) then the latter is less than $n_1/2$ (i.e., $W^{n_1}(f_{1;1}(w_1),\bar{w}_2)$ is good). 

\subsection{Proof of Claim \ref{claim:good2}}
\label{sec:claim:good2}

Consider a good pair $(\byf_1,\byf_2)$.
As the pair $(\byf_1,\byf_2)$ is good, we start by noting that the set 
$(W^{-1})^{n_1}(\byf_1,\byf_2)$ of preimages $(\bxf_1,\bxf_2) \in \cX_1^n \times \cX_2^n$
for which $W^{n_1}(\bxf_1,\bxf_2)=(\byf_1,\byf_2)$ satisfies that $\bxf_1$ must be one of at most $2^{n_1/2}$ vectors in $\cX_1^{n_1}$.
Denote this latter subset of $\cX_1^{n_1}$ by 
${\tt Pre_1}(\byf_1)$.
In addition, for $W^{n_1}(f_{1;1}(w_1),f_{2;1}(w_2,\psi_{2;1}(w_1,w_2)))$ to be equal to $(\byf_{1},\byf_2)$ it must be the case that $w_2$ equals $\byf_2$ or its complement. 
Thus, the number of message pairs $(w_1,w_2)$ such that $W^{n_1}(f_{1;1}(w_1),f_{2;1}(w_2,\psi_{2;1}))=(\byf_1,\byf_2)$ equals $2$ times the number of messages $w_1$ for which $f_{1;1}(w_1) \in {\tt Pre_1}(\byf_1)$. We analyze this latter quantity.

Recall that the codebook $\{x_1^{n_1}(w_1)\}_{w_1 \in [2^{nR_1}]}$ is chosen uniformly at random from the collection of all subsets of size $2^{nR_1}$ of $\mathcal{X}_1^{n_1}$. 
Namely, one can choose the codebook $\{x_1^{n_1}(w_1)\}_{w_1 \in [2^{nR_1}]}$ in an iterative manner, where in iteration $w_1$, $\bxf_1(w_1)$ is chosen uniformly from $\cX_1^{n_1} \setminus \{\bxf_1(w_1')\}_{w_1' <w_1}$. Here, we use the standard order on $[2^{nR_1}]$. 
For any $w_1$, conditioning on the value of $f_{1;1}(w_1')$ for all $w_1'<w_1$, the probability (over the choice of $f_{1;1}(w_1)=\bxf_1(w_1)$) that $f_{1;1}(w_1) \in {\tt Pre_1}(\byf_1)$  is at most 
\begin{equation*}
  \frac{2^{n_1/2}}{4^{n_1}-2^{3n_1/2-\delta n_1}} \leq 2\cdot 2^{-3n_1/2}
\end{equation*}
for sufficiently large $n$. We use the fact that  $n_1=(1-\epsilon)n$, that in each iteration at most $2^{nR_1}= 2^{n_1(3/2-\delta)}$ codewords have been chosen so far, and that our choices are without repetition.
Thus, the probability that there exist $\ell$ messages $\{w_{1,1},\dots, w_{1,\ell}\}$ in $[2^{nR_1}]$ such that for all $i=1,\dots,\ell$ it holds that  
$f_{1;1}(w_{1,i}) \in {\tt Pre_1}(\byf_1)$ is at most 
\begin{eqnarray*}
{{2^{3n_1/2 - \delta n_1}} \choose {\ell}} 2^\ell 2^{-3\ell n_1/2} & \leq &  2^{3\ell n_1/2 - \delta n_1\ell} 2^\ell 2^{-3\ell n_1/2} \\
& = & 2^\ell 2^{- \delta n_1\ell}
\end{eqnarray*}
Setting $\ell=\lceil 3/\delta \rceil+1$, we have, for sufficiently large $n$ (and thus $n_1)$,  that the above probability is at most \ml{$2^{-3n_1}$.
Finally, taking the union bound over possible $\byf_1$, we conclude that with probability at least $1-|\cY_1|^{n_1}\cdot 2^{-3n_1} = 1-  2^{-n_1(3-\log{6})} \geq 1-2^{-0.4n_1}$ }(over $f_{1:1}$) for any good pair $(\byf_1,\byf_2)$, the number of messages $w_1$ for which $f_{1;1}(w_1) \in {\tt Pre_1}(\byf_1)$ is bounded by $\lceil 3/\delta \rceil$. This, in turn, implies that  for any good pair $(\byf_1,\byf_2)$ the list size obtained by the decoder is bounded by $2\lceil 3/\delta \rceil$.

}

\section{Summary}
Consider a network consisting of a discrete MAC
and a CF that has full knowledge of the messages.
In this work, we show that the average-error
sum-capacity of such a network is always a continuous
function of the CF output link capacities; this 
is in contrast to our previous results on maximal-error sum-capacity
\cite{reliability} \mikel{and our current result using only {a constant number of cooperation bits}.}
\mikel{For the average case analysis}, our proof method relies on finding lower and upper bounds
on the sum-capacity and then using a modified
version of a technique developed by Dueck \cite{DueckSC} 
to demonstrate continuity.
\mikel{Our result on maximal-error considers first the maximal-error list-decoding capacity, 
and then reduces list-decoding to unique-decoding. Our result strongly 
relies on the precise functionality of Dueck's MAC \cite{Dueck} and on 
the fact that it has a maximal-error capacity region that differs from 
its average-error capacity region. A deeper understanding of the family of MACs
for which a {constant number of bits (or even a single bit)} of cooperation can affect the maximal-error capacity region 
is left for future research. 

The edge removal problem opens the door to a host of related questions on
general multi-terminal networks. Each question seeks to determine 
whether adding a $\delta$ capacity noiseless channel $e$ to a memoryless network $\cN$ 
results in an average-error capacity region that is not continuous at $\delta=0$.
These questions help us pinpoint whether the addition of asymptotically negligible
cooperation can ever have a non-negligible impact on average-error capacity as it 
can on maximal-error capacity. 
}


\bibliographystyle{IEEEtran}
\bibliography{ref}{}

\end{document}